\documentclass[a4paper]{article}
\pdfoutput=1 

\usepackage{jheppub} 

\usepackage{graphicx,epsfig,wrapfig,amssymb,color,amsmath,bm,breqn}

\usepackage{tikz}
\usetikzlibrary{arrows,decorations.pathreplacing}

\usepackage{subcaption}
\usepackage[T1]{fontenc} 
\usepackage{multirow}
\usepackage{amsmath}
\usepackage{pifont}
\usepackage{float}
\allowdisplaybreaks
\usepackage{soul}
\usepackage[utf8]{inputenc}
\usepackage{braket}

\usepackage{multirow}

\usepackage{lineno} 

\usepackage{hyperref} 
\hypersetup{
    colorlinks,
    citecolor=black,
    filecolor=black,
    linkcolor=black,
    urlcolor=black
}

\newcommand{\Ds}{\displaystyle}

\newcommand{\nn}{\nonumber}

\newcommand{\tr}{\mathrm{tr}}

\newcommand{\sign}{\text{sign}}

\newcommand{\const}{\text{const.}}

\newcommand{\ot}{\leftarrow}

\renewcommand{\(}{\left(}
\renewcommand{\)}{\right)}

\title{Determination of the Collins-Soper Kernel from Lattice QCD}

\author[a]{Maximilian~Schlemmer}
\author[a]{Alexey~Vladimirov}
\author[a]{Christian~Zimmermann}
\author[b]{Michael~Engelhardt}
\author[a]{Andreas~Sch\"afer}

\affiliation[a]{Institut f\"ur Theoretische Physik, Universit\"at Regensburg, D-93040 Regensburg, Germany}
\affiliation[b]{Department of Physics, New Mexico State University, Las Cruces, NM 88003, USA}

\emailAdd{maximilian.schlemmer@ur.de}
\emailAdd{alexey.vladimirov@ur.de}
\emailAdd{christian.zimmermann@ur.de}
\emailAdd{engel@nmsu.edu}
\emailAdd{andreas.schaefer@ur.de}

\abstract{We present lattice results for the non-perturbative Collins-Soper (CS) kernel, which describes the energy-dependence of transverse momentum-dependent parton distributions (TMDs). The CS kernel is extracted from the ratios of first Mellin moments of quasi-TMDs evaluated at different nucleon momenta.The analysis is done with dynamical $N_f=2+1$ clover fermions for the CLS ensemble H101 ($a=0.0854$\,fm, $m_{\pi}=m_K=422$\,MeV). The computed CS kernel is in good agreement with experimental extractions and previous lattice studies.}

\begin{document}
\maketitle
\flushbottom

\section{Introduction}

The last decades witnessed a rapid increase in our understanding of hadron structure, leading to ever more ambitious goals, such as the investigation of multi-dimensional elements of nucleon structure, which are parameterized by functions like generalized parton distributions (GPDs) and transverse momentum-dependent parton distributions (TMDs). These functions contain significantly more information than collinear parton distributions. Determining the multi-dimensional structure is the goal for many modern (such as COMPASS at CERN \cite{Gautheron:2010wva}, RHIC at BNL \cite{Aschenauer:2015eha}, and JLab12 \cite{Dudek:2012vr}) and future (such as the Electron-Ion Collider (EIC) \cite{AbdulKhalek:2021gbh}) experimental facilities. Even so, the precise determination of TMDs and GPDs is exceedingly difficult from experimental data alone. Fortunately, lattice QCD has reached a level of maturity that allows supplementing the experimental data in many ways.
This progress is made possible not only by Exaflop computing becoming a reality but also by the theoretical progress exploring new avenues to extract parton distributions from lattice simulations. As this field is very much in a state of flux and opinions differ strongly on which approach is the most successful, we cite here only a few relevant papers \cite{Braun:2007wv,Ji:2013dva,Ma:2014jla,Radyushkin:2017cyf}. In many cases, lattice QCD and experimental measurements give access to complementary information, making a combined strategy very promising.
In this paper, we discuss one such case, namely the Collins-Soper (CS) kernel \cite{Collins:1981va}, also known as the rapidity anomalous dimension \cite{Chiu:2011qc}.

The CS kernel is a fundamental nonperturbative function that characterizes the QCD vacuum \cite{Vladimirov:2020umg}. It appears as the universal rapidity evolution kernel for TMDs and can be extracted by comparing experimental data at different scales. The CS kernel depends on transverse distance $b$, characterizing the relevant nonlocal parton correlator. The comparison of the most recent extractions made in refs.~\cite{Scimemi:2019cmh,Bacchetta:2019sam,Bertone:2019nxa} demonstrates that presently available data (from Drell-Yan and Semi-Inclusive Deep-Inelastic scattering (SIDIS) reactions) are not sensitive to the value of the CS kernel at $b\gtrsim 2$\,GeV$^{-1}$. The situation will certainly improve with inclusion in the fit of on-going and future measurements, but even the EIC will hardly restrict the CS kernel for $b\gtrsim 3.5$\,GeV$^{-1}$. By contrast, lattice calculations for $b\sim 5$\,GeV$^{-1}$ with reliable uncertainty could be possible within a few years.

Several ways to extract the CS kernel were suggested within the last years \cite{Ebert:2018gzl,Ebert:2019okf,Ji:2019ewn,Vladimirov:2020ofp}. All these suggestions are based on the measurement of quasi-transverse momentum-dependent parton distributions (qTMDs) -- the correlation of two quark fields in a hadron wave function connected by a staple-shaped gauge link positioned in a space-like plane. It has been shown that in the regime of large hadron momentum and large staple length, a qTMD is a convolution of a physical TMD, a perturbative coefficient function, and a soft factor. The latter is an unknown nonperturbative function, depending only on $b$, that cancels (together with other multiplicative factors, particularly the  Wilson line renormalization factors) in the ratio of qTMDs.  The ratios of qTMDs give access to several different pieces of information. In particular, ratios of qTMDs at distinct hadron momenta (with the other parameters being identical) are exclusively sensitive to the CS kernel \cite{Ebert:2018gzl,Vladimirov:2020ofp}. In this work, we evaluate such ratios and extract the CS kernel using the approach suggested in ref.~\cite{Vladimirov:2020ofp}.

One of the fundamental difficulties in the evaluation of quasi-distributions is the necessity to perform a Fourier transformation into momentum-fraction space (introducing the variable $x$). This transformation is notoriously ill-defined, already for one-dimensional observables (for a review of recent developments, see  \cite{Lin:2020rut,Ji:2020ect}), and even more so for qTMDs, where one should respect an entangled hierarchy of several scales. To by-pass this severe complication, we study the first moment of qTMDs (i.e., the integral over $x$). These are much simpler observables that were already studied on the lattice before the formulation of factorization theorems for qTMDs \cite{Musch:2010ka,Musch:2011er,Engelhardt:2015xja,Yoon:2017qzo}. Thus, the methodology of the determination of the first moments of qTMDs is well established. On the theory side, qTMDs are related to TMDs by the factorization theorem, which is valid in the regime of large hadron momentum. The integral over the factorized expression gives us the factorization theorem of the moments, which contains an unknown function that accumulates the information about the $x$-dependence of TMDs. The central aspect of the present approach is the assumption that this function is almost independent of $b$. This is a model assumption. However, all previous studies (theoretical and phenomenological) show that it is a very good approximation. We expect that a systematic error introduced by such an assumption is much smaller than our approach's other uncertainties. However, let us stress that our method has a limited scope of application and is less valuable for determining other properties of TMDs. In contrast, the approaches used in \cite{Zhang:2020dbb,Shanahan:2020zxr} are more general.

As members of the CLS (Coordinated Lattice Simulations) collaboration, we subscribe to its general strategy of focusing on controlling the continuum limit. As the lattice spacing is reduced, standard QCD simulation algorithms suffer from critical slowing down, meaning that an increasing Hybrid Monte Carlo (HMC) simulation time is required to sample the configuration space fully.  Suppose gauge (and fermion) fields fulfill (anti)periodic boundary conditions in all directions. In that case, the simulation will eventually become stuck in a fixed topological sector, and ergodicity is lost, typically for lattice spacing $a<0.05$\,fm. This effect is known as ``topological freezing'' \cite{Luscher:2011kk}. To avoid the ``topological freezing'', the CLS collaboration uses open boundary conditions for lattices with fine spacing. So far, the CLS collaboration has generated about 50 different ensembles, which together allow to reliably control all systematic errors of results extrapolated to the physical point. In this paper, we present results for just one of these ensembles labeled H101. More ensembles and the extrapolation to the physical point will be analyzed in future work. This paper focuses on analyzing the applicability of the method proposed in \cite{Vladimirov:2020ofp} and demonstrates that it is more efficient than the methods used in the refs.~\cite{Zhang:2020dbb} and \cite{Shanahan:2020zxr}.

The paper is partitioned into three main sections. In sec.~\ref{sec:theory} we review the factorization theorem for qTMDs and explain the theoretical setup for our study. The theoretical peculiarities and assumptions of the CS kernel determination from the first moment of qTMDs are described in subsec.~\ref{sec:extraction-cs-kernel}. Sec.~\ref{sec:lattice} is devoted to the details of our lattice analysis. In sec.~\ref{sec:CS} the extraction of the CS kernel is presented. Subsec.~\ref{sec:dermination-of-CS} presents technical details of the extraction, while we discuss systematic uncertainties in subsec.~\ref{sec:uncertanties}. The results are given in subsec.~\ref{sec:discussion}. The paper concludes with section~\ref{sec:conclusion}.

\section{Formalism}
\label{sec:theory}

In this section, we introduce the notation and provide theory details used to determine the CS kernel. For the derivation of the presented expressions cf. ref.~\cite{Vladimirov:2020ofp}.

\subsection{Factorization theorem for qTMDs}

\begin{figure}[b]
\begin{center}
\includegraphics[width=0.6\textwidth]{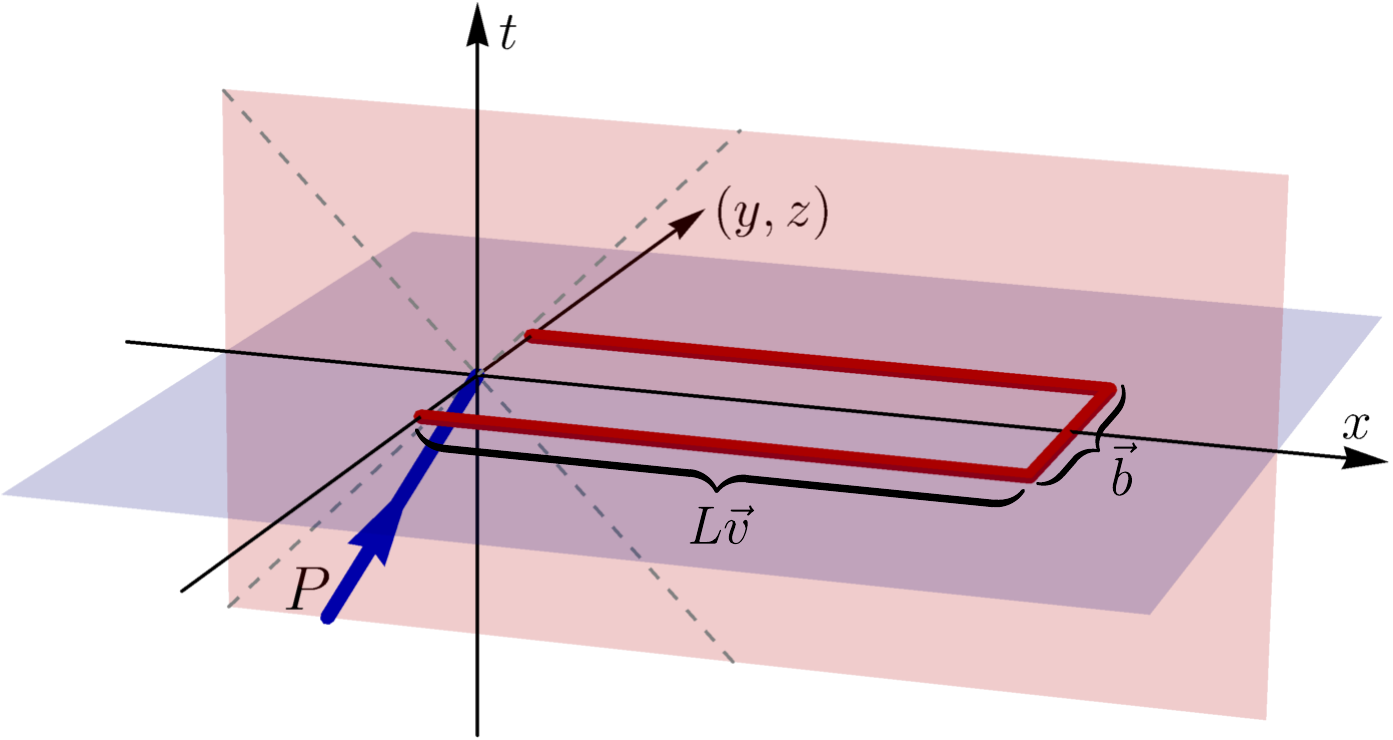}
\end{center}
\caption{\label{fig:staple-geometry} The geometrical configuration of the matrix element (\ref{th:W-def}). The red lines represent the operator, and the blue line represents the hadron momentum.}
\end{figure}

We study the following matrix element
\begin{eqnarray}\label{th:W-def}
&&W_{f \leftarrow h}^{[\Gamma]}(b;\ell, L;v,P,S;\mu)=\frac{1}{2} \langle h(P,S)|\bar{q}_f(b+\ell v)\Gamma\;\mathcal{U}[\mathcal{C}(l, v, b, L)]\;q_f(0)|h(P,S)\rangle
\\\nn && \qquad =
 \frac{1}{2} \langle h(P,S)|\bar{q}_f(b+\ell v)\Gamma[b+\ell v,b+Lv] [b+Lv,Lv] [Lv,0] q_f(0)|h(P,S)\rangle
\end{eqnarray}
where $f$ labels the quark flavor, $\Gamma$ is a Dirac matrix, $[x, y]$ is the straight gauge link between $x$ and $y$, and $|h(P,S)\rangle$ is a single-hadron state with momentum $P$ and spin $S$. The vector $b^\mu$ is orthogonal to the vectors $P^\mu$ and $v^\mu$, $(bP)=(vb)=0$. In addition, the vectors $v^\mu$ and $b^\mu$ lie in an equal-time plane, i.e. they have no time-component and hence $b^2<0$ and $v^2<0$. For definiteness, we fix the normalization by $v^2=-1.$ A visualization of this configuration is given in fig.~\ref{fig:staple-geometry}.

Let us assume that the vectors $P^\mu$ and $v^\mu$ are positioned in the $(t,x)$ plane\footnote{This implies $v^\mu=(0,1,0,0)$. In principle, one can consider the case $v=(0,\vec v)$, where $\vec v$ is not pointing along the $x$ direction. However, in that case the vector $b^\mu$ has only one degree of freedom, due to $(bP)=(vb)=0$ and $b^0=0$. Also, in this case the factorization formula receives additional power corrections $\sim |v\cdot P|-|\vec P|$.}. We introduce light cone coordinates for all four-vectors $a^\mu$, $a^+= (a^0+a^1)/\sqrt{2}$ and $a^-= (a^0-a^1)/\sqrt{2}$. The transverse part of the vector is $a_T^\mu=(0,0,a^2,a^3)$.  The basis vectors corresponding to the $+$ and $-$ components are denoted $\bar n$ and $n$ respectively ($\bar n^2=n^2=0$, $(n\bar n)=1$). The decomposition of the momentum for a nucleon moving in the $x$-direction is
\begin{eqnarray}
P^\mu=P^+ \bar n^\mu+\frac{M^2}{2P^+} n^\mu.
\end{eqnarray}

In the limit of a very energetic hadron and large $L$, the matrix element $W$ can be rewritten in the factorized form
\begin{eqnarray}\label{th:factorization}
W_{f \leftarrow h}^{[\Gamma]}(b;\ell, L;v,P,S;\mu) = \frac{1}{P^+}\int dx e^{ix\ell P^+}\Big|C_H\(\frac{|x|P^+}{\mu}\)\Big|^2
\Phi^{[\Gamma']}_{f\ot h}(x,b;\mu,\zeta)\Psi(b;\mu,\bar \zeta)
+...~,
\end{eqnarray}
where $\Phi$ is a physical TMD distribution, $\Psi$ is a combination of soft-factors \cite{Vladimirov:2020ofp}, and the ellipsis represents the power suppressed correction discussed later. The factorization scales $\mu$, $\zeta$ and $\bar\zeta$ are discussed in the following section. The perturbative coefficient function $C_H$ is known up to next-to-leading order (NLO) \cite{Ebert:2019okf,Vladimirov:2020ofp}. It reads
\begin{eqnarray}\label{th:coeff-function}
\Big|C_H\(\frac{p}{\mu}\)\Big|^2=1+C_F\frac{\alpha_s(\mu)}{4\pi}\(-\ln^2\(\frac{4p^2}{\mu^2}\)+2\ln\(\frac{4p^2}{\mu^2}\)-4+\frac{\pi^2}{6}\)+\mathcal{O}(\alpha_s^2).
\end{eqnarray}
Note that the coefficient function is strictly universal and independent of $\Gamma$. The expression on the RHS of (\ref{th:factorization}) contains
\begin{eqnarray}\label{th:gamma-prime}
\Gamma'=\frac{\gamma^+\gamma^-\Gamma \gamma^-\gamma^+}{4}.
\end{eqnarray}
The operation on the RHS of (\ref{th:gamma-prime}) selects the TMD distributions of leading twist, whereas higher-twist distributions are nullified and appear as part of the power corrections in (\ref{th:factorization}).

The  factorization theorem (\ref{th:factorization}) is valid in the parameter range defined by
\begin{eqnarray}\label{th:fac-limit}
\frac{P^-}{P^+}\ll 1,\qquad \frac{1}{|b|P^+}\ll 1,\qquad \frac{|b|}{L}\ll 1,\qquad \frac{\ell}{L}\ll 1, \qquad \ell \Lambda_{\text{QCD}}\ll 1,
\end{eqnarray}
where $\Lambda_{QCD}$ is the characteristic low-energy scale of QCD. Essentially, one needs large hadron momentum,  large longitudinal size of the contour $L$, and fixed transverse size $b$. The parameter $\ell$ must not be too large to guarantee the $\ell$-independence of the function $\Psi$.

The presence of unknown nonperturbative factors prevents the direct determination of TMDs from the lattice matrix element (\ref{th:W-def}). To eliminate the unknown and singular factor $\Psi$ we consider a ratio of $W$'s evaluated at the same value of $b$ but for different momenta. In this case, the functions $\Psi$ cancel, and the result is expressed entirely in terms of the physical TMDs. For the extraction of the CS kernel, we use the ratio
\begin{eqnarray}\label{th:ratio0}
R_{f\ot h}^{[\Gamma]}(P_1,P_2;b;\ell, L;v,S)=\frac{P_1^+ W_{f \leftarrow h}^{[\Gamma]}(b;\ell, L;v,P_1,S;\mu)}{P_2^+ W_{f \leftarrow h}^{[\Gamma]}(b;\ell, L;v,P_2,S;\mu)}.
\end{eqnarray}
Substituting the expression (\ref{th:factorization}) we obtain
\begin{eqnarray}\label{th:ratio1}
R_{f\ot h}^{[\Gamma]}(P_1,P_2;b;\ell, L;v,S)=\frac{\Ds\int dx_1 e^{ix_1\ell P^+_1}|C_H(|x_1|P_1^+/\mu)|^2 \Phi^{[\Gamma']}_{f\ot h}(x_1,b;\mu,\zeta_1)}{\Ds
\int dx_2 e^{ix_2\ell P^+_2}|C_H(|x_2|P_2^+/\mu)|^2 \Phi^{[\Gamma']}_{f\ot h}(x_2,b;\mu,\zeta_2)}+...~,
\end{eqnarray}
where the ellipsis denotes power suppressed terms. To cancel the factors $\Psi$ we took $\bar \zeta_1=\bar \zeta_2=\bar \zeta$, which according to (\ref{th:zeta*zeta}) implies $\zeta_i=(2x_iP_i^+v^-)^2\mu^2/\bar \zeta$. Note that the ratio (\ref{th:ratio1}) is independent of $\mu$.
The factorization theorems derived in refs.\cite{Ebert:2019okf,Ji:2019ewn} are equivalent to (\ref{th:factorization}). The only difference is that the authors of refs.~\cite{Ebert:2019okf,Ji:2019ewn} consider the qTMD together with a certain soft factor $S_{\text{bent}}$, which is equivalent to the division of (\ref{th:factorization}) by $S_{\text{bent}}$. The factor $S_{\text{bent}}$ is constructed such that it cancels $\Psi$ in the perturbative regime. Performing the inverse Fourier transformation, one determines the physical TMD distribution from the combination of qTMD and $S_{\text{bent}}$. This approach gives access to the full TMD. However, it contains several fundamental complications. The two main complications are -- the absence of proof that $\Psi/S_{\text{bent}}=1$ nonperturbatively and the fact that the inverse Fourier integration incorporates values of $\ell$ that violate the requirements of the factorization theorem (\ref{th:fac-limit}). We expect these complications to be resolved in the future. For this work, the dependence on $x$ is irrelevant since we are only interested in determining the CS kernel. We extract the CS-kernel from the ratios (\ref{th:ratio1}) evaluated at fixed $\ell$.

\subsection{CS kernel and evolution of TMD distributions}

The factorization expression (\ref{th:factorization}) contains a number of scales, which are combined into the variables $\mu$, $\zeta$ and $\bar \zeta$. The ultraviolet renormalization of $W$ gives the overall dependence on $\mu$ \cite{Chetyrkin:2003vi},
\begin{eqnarray}
\mu^2\frac{d W}{d\mu^2}=\(-3C_F \frac{\alpha_s(\mu)}{4\pi}+O(\alpha_s^2)\)W,
\end{eqnarray}
where we omitted the arguments of $W$ for brevity. The factorization of collinear singularities introduces further dependence on $\mu$, which cancels out between the coefficient functions and the functions $\Phi$ and $\Psi$. The scales $\zeta$ and $\bar\zeta$  result from the factorization of rapidity divergences \cite{Vladimirov:2017ksc}. These scales satisfy the relation
\begin{eqnarray}\label{th:zeta*zeta}
\zeta \bar \zeta=(2xP^+ v^-)^2\mu^2.
\end{eqnarray}
Let us emphasize that the scale $\mu$ is present on the RHS of (\ref{th:zeta*zeta}). In  ordinary TMD factorization \cite{Vladimirov:2017ksc,Collins:2011zzd,GarciaEchevarria:2011rb} one has the relation $\zeta\bar \zeta=(2p_1^+p_2^-)$, where $p_i$ are the momenta of the colliding hadrons. In the factorization of $W$ leading to (\ref{th:zeta*zeta}) the second hadron is absent and replaced by Wilson lines. Therefore its momentum does not appear in (\ref{th:zeta*zeta}).

The dependence on $\mu$ and $\zeta$ of the TMD $\Phi$ is given by a pair of equations,
\begin{eqnarray}\label{th:TMD-UV}
\frac{d \ln \Phi^{[\Gamma]}_{f\ot h}(x,b;\mu,\zeta)}{d\ln \mu^2}&=&\frac{\gamma^f_F(\mu,\zeta)}{2},
\\\label{th:TMD-RAP}
\frac{d \ln \Phi^{[\Gamma]}_{f\ot h}(x,b;\mu,\zeta)}{d\ln \zeta}&=&\frac{K^f(b,\mu)}{2},
\end{eqnarray}
where $\gamma_F$ is the ultraviolet anomalous dimension, and $K$ is the CS kernel. The anomalous dimension and the CS kernel depend only on the color representation $F$ and quark flavor $f$. Since in the present work, we deal only with quark TMDs, the label $f$ for $\gamma_F^f$ and $K^f$ is not necessary, and we omit it in the following. Note that the CS kernel itself, which should be independent of the hadron type, depends on the employed momenta $P_1$ and $P_2$ only via the scale $\mu$. However, $P_1$ and $P_2$ do influence the systematic error due to power corrections discussed in subsec.~\ref{sec:uncertanties}.

The integrability condition for the system (\ref{th:TMD-UV}, \ref{th:TMD-RAP}) yields
\begin{eqnarray}\label{th:integrability}
\frac{d K(\mu,b)}{d\ln \mu^2}=\frac{d\gamma_F(\mu,\zeta)}{d\ln\zeta}=-\Gamma_{\text{cusp}}(\mu),
\end{eqnarray}
where $\Gamma_{\text{cusp}}$ is the cusp anomalous dimension of light-like Wilson lines. The expression for $\gamma_F$ is known up to N$^3$LO. The CS kernel is a generically nonperturbative function, but for small values of $b$ it can be computed by means of the weak field approximation. The leading term is
\begin{eqnarray}\label{th:RAD-LO}
K(b,\mu)=-C_F\frac{\alpha_s}{\pi}\mathbf{L}+O(\alpha_s \mathbf{L}^2)+O(b^2),
\end{eqnarray}
where $\mathbf{L}=\ln(|b^2|\mu^2/4 e^{-2\gamma_E})$ with $\gamma_E=0.5772...$ being the Euler constant, and $C_F=4/3$. The N$^3$LO expression is derived in ref. \cite{Vladimirov:2016dll}, and the leading power correction in ref. \cite{Vladimirov:2020umg}. In practice, it is convenient to resum the logarithms of $(b\mu)$, which significantly improves the perturbative convergence of the series \cite{Echevarria:2012pw}. In this case the leading term is
\begin{eqnarray}\label{th:RAD-resum}
K(b,\mu)=\frac{4C_F}{\beta_0}\ln\(1-\beta_0 \frac{\alpha_s}{4\pi}\mathbf{L}\)+O(\alpha_s)+O(b^2),
\end{eqnarray}
where $\beta_0$ is the leading order QCD beta function $\beta_0=\frac{11}{3}N_c-\frac{2}{3}N_f$.

The solution of the system (\ref{th:TMD-UV},\ref{th:TMD-RAP}) is
\begin{eqnarray}
\Phi^{[\Gamma]}_{f\ot h}(x,b;\mu,\zeta)=\exp\Big[
\int_P \(\gamma_F(\mu,\zeta)\frac{d\mu}{\mu}+\frac{K(b,\mu)}{2}\frac{d\zeta}{\zeta}\)\Big]
\Phi^{[\Gamma]}_{f\ot h}(x,b;\mu_0,\zeta_0),
\end{eqnarray}
where $P$ is an arbitrary path connecting the points $(\mu,\zeta)$ and $(\mu_0,\zeta_0)$~\cite{Scimemi:2018xaf}. Path-independence is guaranteed by the integrability condition (\ref{th:integrability}). For our purposes, it will be convenient to evolve TMDs along the path of constant $\mu$. Choosing the straight path we obtain
\begin{eqnarray}\label{th:TMD-fixedmu}
\Phi^{[\Gamma]}_{f\ot h}(x,b;\mu,\zeta)=\(\frac{\zeta}{\zeta_0}\)^{K(b,\mu)/2}\Phi^{[\Gamma]}_{f\ot h}(x,b;\mu,\zeta_0).
\end{eqnarray}

\subsection{Extraction of the CS kernel from ratios at $\ell=0$}
\label{sec:extraction-cs-kernel}

\begin{figure}[t]
\begin{center}
\includegraphics[width=0.45\textwidth]{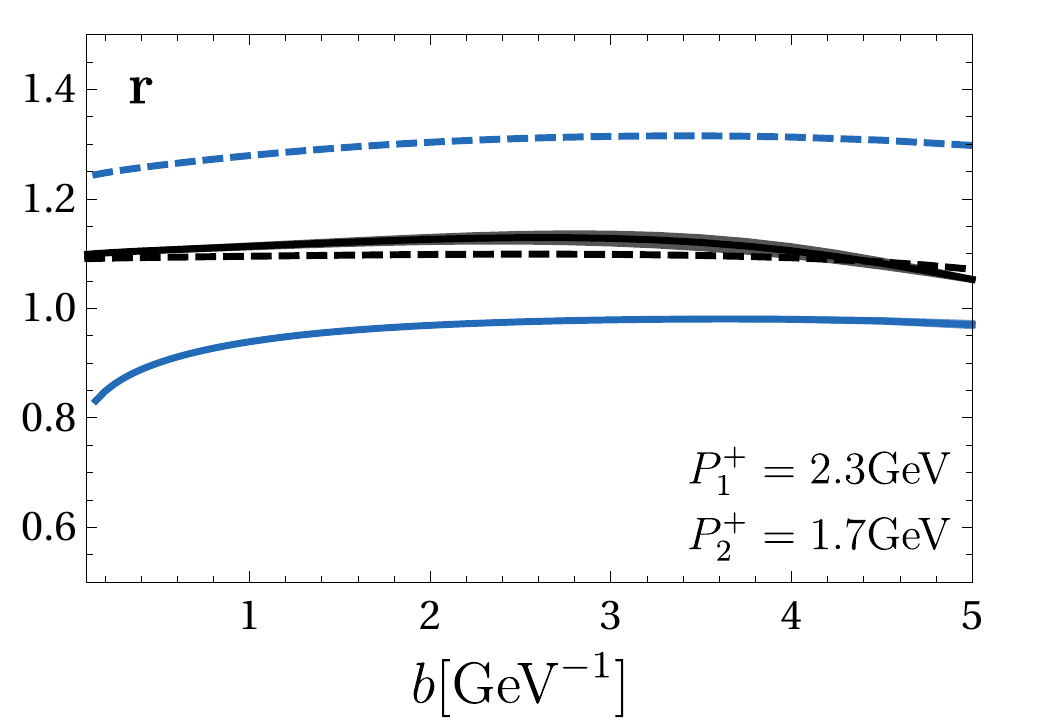}
\includegraphics[width=0.45\textwidth]{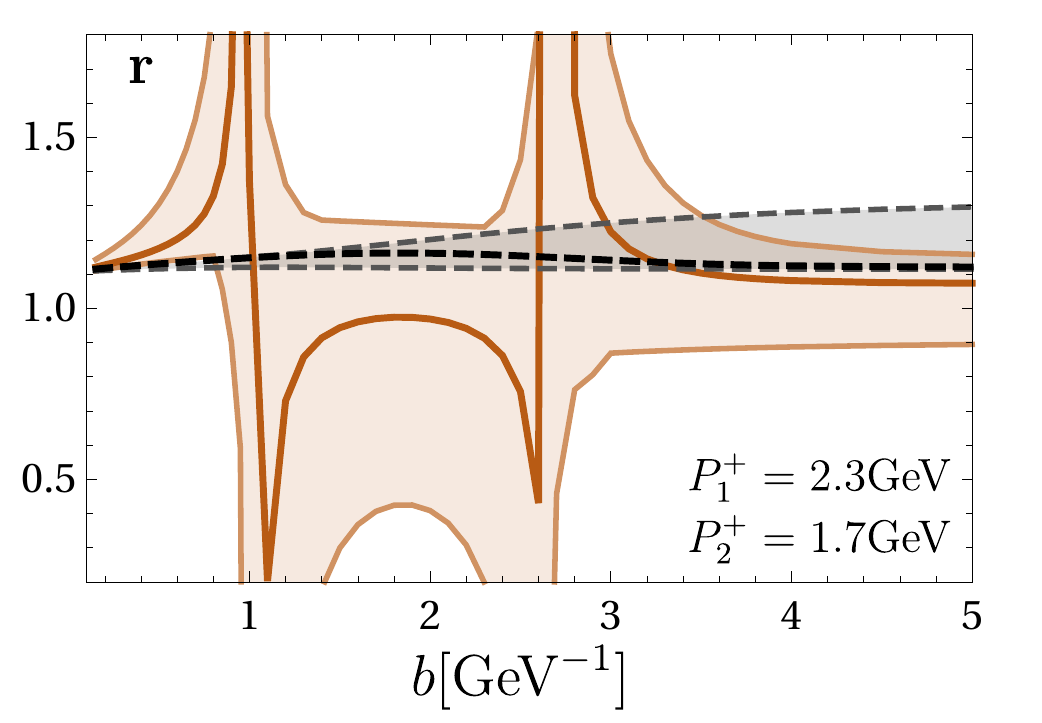}
\end{center}
\caption{\label{fig:2} (left) The factor $\mathbf{r}$ computed for d-quark unpolarized TMDs in the proton (blue) and pion (black), using the phenomenological extractions \cite{Scimemi:2019cmh,Vladimirov:2019bfa}. The solid line (with an uncertainty band, due to the extraction uncertainties) is the result for the direct evaluation (\ref{th:r}). The dashed line is the result of the perturbative computation (\ref{th:r-pert}) with the function $\mathbf{M}$ shown in fig.~\ref{fig:1}.\\
(right)
The factor $\mathbf{r}$ was computed for the d-quark Sivers function in the proton, using the phenomenological extraction \cite{Bury:2020vhj}. The orange line (with an uncertainty band due to the extraction uncertainty) presents the direct evaluation (\ref{th:r}). The black line is the result of perturbative computation (\ref{th:r-pert}), and the function $\mathbf{M}$ is shown in fig.~\ref{fig:1}. The Sivers function is not sign-definite, which produces diverging uncertainty bands for the ratio.}
\end{figure}

In our lattice simulation we evaluate $W$ at $\ell=0$. In this case, the expression (\ref{th:ratio1}) simplifies to
\begin{eqnarray}\label{th:ratio2}
R_{f\ot h}^{[\Gamma]}(P_1,P_2;b)=\frac{\Ds\int dx_1 |C_H(|x_1|P_1^+/\mu)|^2 \Phi^{[\Gamma']}_{f\ot h}(x_1,b;\mu,\zeta_1)}{\Ds
\int dx_2|C_H(|x_2|P_2^+/\mu)|^2 \Phi^{[\Gamma']}_{f\ot h}(x_2,b;\mu,\zeta_2)}+...~,
\end{eqnarray}
where we also drop unimportant variables $(L,v,S)$ from the argument for brevity. To explicitly extract the CS kernel we evolve both TMDs in $\zeta$ to the point $\zeta_0$ using (\ref{th:TMD-fixedmu}), and obtain
\begin{eqnarray}\label{th:ratio-cs-kernel}
R_{f\ot h}^{[\Gamma]}(P_1,P_2;b)=\(\frac{P_1^+}{P_2^+}\)^{K(b,\mu)}\mathbf{r}^{[\Gamma]}_{f\ot h}(b,\mu;P_1,P_2)+...~,
\end{eqnarray}
where
\begin{eqnarray}\label{th:r}
\mathbf{r}^{[\Gamma]}_{f\ot h}(b,\mu;P_1,P_2)=\frac{\Ds\int dx_1 |x_1|^{K(b,\mu)} |C_H(|x_1|P_1^+/\mu)|^2 \Phi^{[\Gamma']}_{f\ot h}(x_1,b;\mu,\zeta_0)}{\Ds
\int dx_2 |x_2|^{K(b,\mu)} |C_H(|x_2|P_2^+/\mu)|^2 \Phi^{[\Gamma']}_{f\ot h}(x_2,b;\mu,\zeta_0)}.
\end{eqnarray}
In general, the function $\mathbf{r}^{[\Gamma]}_{f \ot h}$ has complicated properties. For example, its numerator and denominator have potential problems with convergence at $x\to 0$, see the discussion in ref.~\cite{Vladimirov:2020ofp}. Also, the function $\mathbf{r}^{[\Gamma]}_{f \ot h}$ depends on $\mu$. To simplify it and reveal its dependence on $\mu$, we expand $\mathbf{r}^{[\Gamma]}_{f \ot h}$ in the limit of small $\alpha_s$.
The NLO perturbative expansion for this function is
\begin{eqnarray}\label{th:r-pert}
&& \mathbf{r}^{[\Gamma]}_{f\ot h}(b,\mu;P_1,P_2)
=1
\\\nn &&
\qquad +4C_F\frac{\alpha_s(\mu)}{4\pi}\ln\(\frac{P_1^+}{P_2^+}\)\Big[1-\ln\(\frac{4P_1^+P_2^+|v^-|^2}{\mu^2}\)-2 \mathbf{M}^{[\Gamma]}_{f\ot h}(b,\mu)\Big]
+O(\alpha_s^2),
\end{eqnarray}
where
\begin{eqnarray}\label{th:M-def}
\mathbf{M}^{[\Gamma]}_{f\ot h}(b,\mu)=\frac{\Ds\int dx_1 \ln|x_1|x_1^{K(b,\mu)}\Phi^{[\Gamma]}_{f\ot h}(x_1,b;\mu,\zeta_0)}{\Ds\int dx_2 x_2^{K(b,\mu)}\Phi^{[\Gamma]}_{f\ot h}(x_2,b;\mu,\zeta_0)}.
\end{eqnarray}
In the following we use this expression as our approximation for $\mathbf{r}^{[\Gamma]}_{f \ot h}$, assuming  $\mathbf{M}^{[\Gamma]}_{f \ot h}$ to be a constant in $b$. The scale $\mu$ is selected to nullify the logarithm in the square brackets of (\ref{th:r-pert})
\begin{eqnarray}\label{th:mu}
\mu=\sqrt{2P^+_1P^+_2},
\end{eqnarray}
where we used that $|v^-|=1/\sqrt{2}$.

\begin{figure}[t]
\begin{center}
\includegraphics[width=0.6\textwidth]{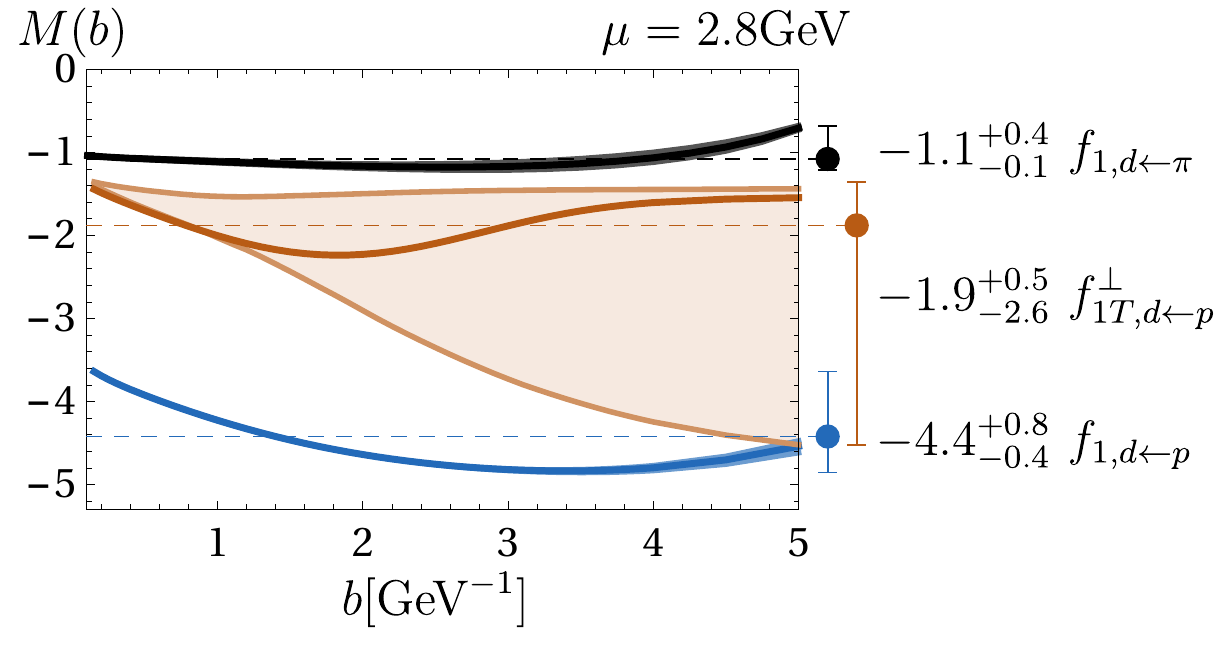}
\end{center}
\caption{\label{fig:1}  The function $\mathbf{M}^{[\Gamma]}_{f \ot h}$ (\ref{th:M-def}) computed using fits for the d-quark unpolarized TMDs in the proton (blue) and in the pion (black), and from the d-quark Sivers function (brown) in refs.\cite{Scimemi:2019cmh,Vladimirov:2019bfa,Bury:2020vhj}, with uncertainty bands. The numbers to the right show the mean and the error for each case, where the error is given by the maximal deviation. Note that for the unpolarized TMD $f_1$ there is more precise experimental data available than for the Sivers function $f_{1T}$ and thus $\mathbf{M}^{[f_1]}_{f \ot h }(b)$ can be determined with higher precision than $\mathbf{M}^{[f_{1T}]}_{f \ot h }(b)$.}
\end{figure}

The central point of our approach is the assumption that the function $\mathbf{r}^{[\Gamma]}_{f \ot h}$ is almost independent of $b$. Such a behavior is expected, because a different behavior of denominator and numerator in (\ref{th:r}) can only come from the $\ln x$ term present in $C_H$. Therefore, any essential deviation from the constant behavior implies a significant change of the $x$-profile between different values of $b$, which is unlikely. In fig.~\ref{fig:2} we present examples of the function $\mathbf{r}^{[\Gamma]}_{f \ot h}$ for different TMDs computed using phenomenological extractions \cite{Scimemi:2019cmh,Vladimirov:2019bfa,Bury:2020vhj}. The deviation of $\mathbf{r}^{[\Gamma]}_{f \ot h}$ from its mean value for $b>1$\,GeV$^{-1}$ is (+1.2\%,-3.5\%) for the unpolarized TMD in the proton, and (+2\%,-5\%) for the unpolarized TMD in the pion. Let us mention that in the case of the Sivers function, the integrals are not sign-definite, and thus the ratios (\ref{th:r}) are singular at certain points (see fig.~\ref{fig:2}(right)). This results in the enormous error band for $\mathbf{r}_{\text{Sivers}}$, and indicates possible issues in an application of the suggested method in this case. A phenomenological estimate of $\mathbf{M}^{[\Gamma]}_{f \ot h}$ is presented in fig.~\ref{fig:1}. The computation of $\mathbf{r}^{[\Gamma]}_{f \ot h}$ using (\ref{th:r-pert}) based on this phenomenological $\mathbf{M}^{[\Gamma]}_{f \ot h}$ is given by the dashed lines in fig.~\ref{fig:2}.

The observation that $\mathbf{r}^{[\Gamma]}_{f \ot h}$ (or equivalently $\mathbf{M}^{[\Gamma]}_{f \ot h}$) is approximately independent of $b$ is central for our method of determining the CS kernel. Assuming $\mathbf{r}^{[\Gamma]}_{f \ot h}$ to be constant we determine the CS kernel by
\begin{eqnarray}\label{th:cs-kernel-explicit}
K(b,\mu)=\frac{\ln(R)-\ln(\mathbf{r})}{\ln(P_1^+/P_2^+)},
\end{eqnarray}
where we omit the arguments of $R$ and $\mathbf{r}$. To determine the constant $\mathbf{M}^{[\Gamma]}_{f \ot h}$ we normalize $K$ computed on the lattice to the perturbation theory values with $b$ in the regime where perturbation theory and the factorization formula are applicable. That requires simultaneously $b\gg 1/P^+$ and $b\ll 1/\Lambda_{\text{QCD}}$. We select $b\sim 1\,\textnormal{GeV}^{-1}$ for which the perturbative series converges well \cite{Ebert:2018gzl,Echevarria:2012pw}. For normalization we use the N$^3$LO value in the resummed form \cite{Vladimirov:2016dll,Scimemi:2019cmh}.
The normalization procedure introduces a correlated systematic uncertainty. Assuming that $\mathbf{r}$ ($\mathbf{M}$) has variance $\delta \mathbf{r}$ ($\delta \mathbf{M}$) we compute the variance for $K$ as
\begin{eqnarray}\label{th:deltaK1}
\delta K=\frac{\delta \mathbf{r}}{\mathbf{r}}\frac{1}{\ln(P_1^+/P_2^+)}=8C_F\frac{\alpha_s}{4\pi}\frac{\delta \mathbf{M}}{\mathbf{r}}.
\end{eqnarray}
Considering the phenomenological extractions shown in figs.\ref{fig:2}, we obtain $\delta K\sim 0.15$ for the unpolarized proton case, and $\delta K\sim 0.06$ for the unpolarized pion case. In the case of the Sivers function, this uncertainty is larger, $\delta K\sim 0.25$. Therefore, using this approach, we can estimate the CS kernel from $\ell=0$ matrix elements with a systematic uncertainty of about $\pm 0.1$. Let us emphasize that the method described here does not depend on the error in the determination of absolute values of $\mathbf{r}$, but only on the error of the deviation of $\mathbf{r}(b)$ from the constant.

The analysis presented in this section is valid only for $\Gamma$-matrices corresponding to leading twist TMDs, for which $\Gamma'=\Gamma$ (\ref{th:gamma-prime}). For leading twist TMDs, the CS kernel is independent of $\Gamma$. This fact can be exploited for a cross-check of our computation. There are three $\Gamma$-matrices of leading twist TMDs
\begin{eqnarray}\label{def:leading-twist}
\Gamma=\{\gamma^+,\gamma^+\gamma^5,i\sigma^{\alpha+}\gamma^5\},
\end{eqnarray}
where the index $\alpha$ is transverse. For these cases, the matrix elements $\Phi^{[\Gamma]}$ are parameterized in terms of the leading twist TMDs according to \cite{Mulders:1995dh,Scimemi:2018mmi}
\begin{eqnarray}\label{def:vector}
\Phi^{[\gamma^+]}_{f\ot h}(x,b)&=&f_1(x,b)+i\epsilon^{\mu\nu}_Tb_{\mu}s_{T\nu}Mf_{1T}^\perp(x,b),
\\\label{def:axial}
\Phi^{[\gamma^+\gamma^5]}_{f\ot h}(x,b)&=&\lambda g_{1L}(x,b)+ib_{\mu}s_{T}^\mu M g_{1T}(x,b),
\\\label{def:tensor}
\Phi^{[i\sigma^{\alpha+}\gamma^5]}_{f\ot h}(x,b)&=&s_T^\alpha h_1(x,b)-i\lambda b^\alpha M h_{1L}^\perp(x,b)
\\\nn &&
+i\epsilon^{\alpha\mu}_T b_{\mu} h_1^\perp(x,b)+\frac{M^2b^2}{2}\(\frac{g_T^{\alpha\mu}}{2}-\frac{b^\alpha b^\mu}{b^2}\)s_{T\mu} h_{1T}^\perp(x,b),
\end{eqnarray}
where $\epsilon^{\mu\nu}_T$ and $g_T^{\mu\nu}$ are the transverse parts of the Levi-Civita and the metric tensor ($\epsilon^{23}_T=-\epsilon^{32}_T=1$, if the transverse plane is the $(y,z)$ plane), $s_T^\mu$ and $\lambda$ are the transverse and longitudinal components of the nucleon's spin vector, and $M$ is the mass of the nucleon. Considering different polarization states of the hadron, we can access different TMDs independently.

The adopted method's main advantage is that lattice computations of $R(\ell=0)$ are essentially simpler than the evaluation of qTMDs with $x$-dependence. In our case, we do not need to account for $\ell$-dependent renormalization factors and power corrections induced by $\ell$. We see this as an advantage of our analysis. In any case, the CS kernel extraction might be strongly affected by power corrections, which (as we expect) is a common property of all determinations of the CS kernel from the lattice. Better control of power corrections is vital for continued progress. At present, too little is known about these to predict which precision can be ultimately reached. The only point that seems inevitable is that further progress will require substantial work on all sides -- perturbative QCD, lattice QCD, and experiment.

\section{Evaluation of TMD matrix elements on the lattice}
\label{sec:lattice}

In this section we provide details of the computation of the matrix elements $W^{[\Gamma]}$ (\ref{th:W-def}) on the lattice.

\subsection{Computation of the TMD matrix elements}
\label{sec:computation1}

In the following, we consider Euclidean spacetime, and utilize the following notation for any four-vector, $a = (\vec{a}, a^4)$, i.e.\ $\vec{a}$ represents the spatial vector components, whereas $a^4$ is the Euclidean time component.
The TMD matrix element can be accessed on an Euclidean lattice by evaluating an extended three point function $C_{3\mathrm{pt}}$. It is defined as:
\begin{align}
\label{eq:3pt-def}
C^{\Gamma}_{3\mathrm{pt}}(\vec{P},S,\mathcal{C},t,\tau) := \left\langle \tr \left\{ \Gamma_S \mathcal{P}(\vec{P},t)\ J^\Gamma(\mathcal{C},\tau)\ \overline{\mathcal{P}}(\vec{P},0) \right\} \right\rangle\,,
\end{align}
where the angle brackets indicate the sum over all gauge configurations of the employed gauge ensemble and the trace is taken with respect to spinor indices. The operators $\mathcal{P}$ and $\overline{\mathcal{P}}$ are the proton interpolators
\begin{align}
\label{eq:proton-interpolators}
\mathcal{P}(\vec{k},t) &:= a^3 \sum_{\vec{x}} e^{-i\vec{x}\vec{k}} \left. \epsilon_{abc} u_a(x) \left[ u_b^T(x) C\gamma_5 \frac{1 + \gamma_4}{2} d_c(x) \right] \right|_{x^4=  t}\,, \nonumber \\
\overline{\mathcal{P}}(\vec{k},t) &:= a^3 \sum_{\vec{x}} e^{i\vec{x}\vec{k}} \left. \epsilon_{abc} \left[ \bar{u}_a^T(x) C\gamma_5 \frac{1 + \gamma_4}{2} \bar{d}_b(x) \right] \bar{u}_c(x)  \right|_{x^4=t}\,.
\end{align}
These annihilate or create a tri-quark state sharing the same quantum numbers as the proton, which is implemented in the Chroma software system~\cite{EDWARDS2005832}. The spinor matrix $\Gamma_S$ projects out the state with positive parity (at rest) and the desired proton spin $S$, see (\ref{eq:lattice-spin-proj}). The expression \eqref{eq:3pt-def} is invariant under shifts in time, assuming infinite or periodic time extensions. In our simulations, we use a lattice with open boundaries in the time direction. For that reason we shall use the source time slice $t_{\mathrm{src}} \gg 0$.

In the current case, the operator $J^\Gamma$ is represented by a non-local gauge-invariant quark bilinear (\ref{th:W-def}),
\begin{align}
\label{eq:def-quark-bilinear}
J^\Gamma(\mathcal{C},\tau) := \bar{q}(b+\ell v,\tau) \Gamma \mathcal{U}[\mathcal{C}(\ell, v, b, L)] q(0,\tau)\,.
\end{align}
The vectors $b$ and $v$ have vanishing time components and can be directly translated to Minkowski spacetime. Hence, the quark fields are positioned at the same Euclidean time slice $\tau$. The operator $J^\Gamma$ contains the staple-shaped gauge link $\mathcal{U}[\mathcal{C}]$ defined in \eqref{th:W-def}. The lattice analog of $\mathcal{U}$ can be constructed with the elementary gauge links $U_\mu(x)$. In our simulations, we use Wilson lines $[x,x+y]$ with two possible orientations, $y=\hat{\mu}N$ and $y=(\hat{\mu}+\hat{\nu})N$. Here, the variables $\hat{\mu}$ and $\hat{\nu}$ are the lattice unit vectors pointing in different directions and $N$ is an integer number. The explicit construction of $[x,x+y]$ in terms of elementary links is
\begin{align}
\label{eq:lattice-gauge-links-straight}
[x,x+y] &= \prod_{n=0}^{N-1} U_\mu (x+n\hat{\mu}) &~~~&\text{for}~ &y &= \hat{\mu}N
\,, \\
\label{eq:lattice-gauge-links-step}
[x,x+y] &= \prod_{n=0}^{N-1} U_\mu (x+n(\hat{\mu}+\hat{\nu}))\ U_\nu(x+(n+1)\hat{\mu}+n\hat{\nu}) &~~~&\text{for}~ &y &= (\hat{\mu}+\hat{\nu})N\,.
\end{align}
Notice that in the latter case, there are two nonequivalent possibilities to implement the link path. The second is obtained by interchanging $\hat \mu$ and $\hat \nu$. We perform our calculations using both versions.

In order to relate the three-point function $C_{3\mathrm{pt}}$ to the TMD matrix element $W_{f\leftarrow h}^{[\Gamma]}$ we calculate the following quantity:

\begin{align}
\label{eq:def-lattice-ratio}
\left. 2\sqrt{m^2+\vec{P}^2}\frac{C_{3\mathrm{pt}}(P,S,\mathcal{C},t,\tau)}{C_{2\mathrm{pt}}(P,S,t)} \right|_{0\ll \tau \ll t} = \frac{\sum_{rs} \bar{u}(r,P) \Gamma_S u(s,P) \bra{h(P,s)} J^\Gamma(\mathcal{C},\tau) \ket{h(P,r)}}{\sum_s \bar{u}(s,P) \Gamma_S u(s,P)} \,,
\end{align}
where the r.h.s.\ can be identified with $W_{f\leftarrow h}^{[\Gamma]}$ if $\Gamma_S$ is chosen accordingly. In equation \eqref{eq:def-lattice-ratio} the Euclidean time separations between source, insertion, and sink should be large. In this limit, the excited proton states, which also overlap with the proton interpolators, are exponentially suppressed. The ratio with the two-point function is constructed using
\begin{align}
\label{eq:2pt-def}
C_{2\mathrm{pt}}(\vec{P},S,t) := \left\langle \tr \left\{ \Gamma_S \mathcal{P}(\vec{P},t)\ \overline{\mathcal{P}}(\vec{P},0) \right\} \right\rangle\,.
\end{align}
This weighting is required for the correct normalization.

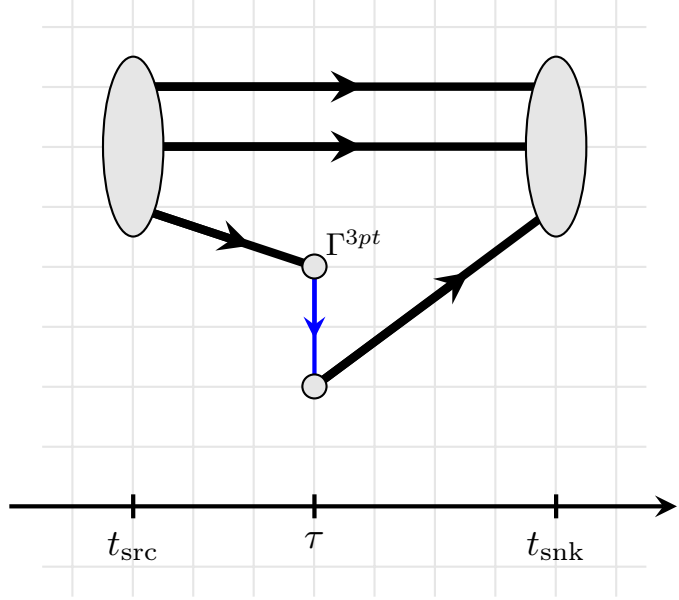
\begin{figure}
\begin{center}
\resizebox{0.6\textwidth}{!}{\begin{tikzpicture}[x=0.4cm,y=0.4cm,z=0.1cm,>=stealth]
    \draw[step=1.0,gray!20,thin,xshift=0.5,yshift=0.5] (0.5,0.5) grid (10.5,10.5);

    \draw[->,ultra thick] (2.05,9.05) -- (5.85,9.05);
    \draw[ultra thick] (2.05,9.05) -- (9.05,9.05);
    \draw[->,ultra thick] (2.05,8.05) -- (5.85,8.05);
    \draw[ultra thick] (2.05,8.05) -- (9.05,8.05);

    \draw[->,ultra thick] (2.05,7.05) -- +(1.95,-0.65);
    \draw[ultra thick] (2.05,7.05) -- (5.05,6.05);
    \draw[->,thick,blue] (5.05,6.05) -- +(0,-1.2);
    \draw[thick,blue] (5.05,6.05) -- (5.05,4.05);
    \draw[->,ultra thick] (5.05,4.05) -- +(2.55,1.90);
    \draw[ultra thick] (5.05,4.05) -- (9.05,7.05);

    \draw[black,fill=gray!20] (2.05,8.05) ellipse (0.5 and 1.5);
    ß
    \draw[black,fill=gray!20] (9.05,8.05) ellipse (0.5 and 1.5);

    \draw[black,fill=gray!20] (5.05,6.05) circle (0.2) node[anchor=south west,scale=0.6] {$\Gamma^{3pt}$};
    \draw[black,fill=gray!20] (5.05,4.05) circle (0.2);

    \draw[->,thick] (0.,2.05) -- (11.05,2.05);
    \draw[thick] (2.05,2.25) -- (2.05,1.85) node[anchor=north,scale=0.7] {$t_{\textnormal{src}}$};
    \draw[thick] (5.05,2.25) -- (5.05,1.85) node[anchor=north,scale=0.7] {$\tau$};
    \draw[thick] (9.05,2.25) -- (9.05,1.85) node[anchor=north,scale=0.7] {$t_{\textnormal{snk}}$};
\end{tikzpicture}}
\end{center}
\caption{\label{fig:nucleon-three-pt} Lattice nucleon three point function with the source nucleon located at $t_{\textnormal{src}}$ and the sink nucleon at $t_{\textnormal{snk}}$ separated by the distance $t = t_{\textnormal{snk}} - t_{\textnormal{src}}$. The blue line is the Wilson line with the shape shown in fig.~\ref{fig:staple-geometry}. }
\end{figure}

The evaluation of the fermionic integral implicit in \eqref{eq:3pt-def} leads to several Wick contractions, of which we distinguish two kinds. These are referred to as "connected" and "disconnected" graphs, where the latter involves a fermion loop, including the operator $J^\Gamma$. In this study, we restrict ourselves to the non-singlet quark contributions (particularly, $u-d$). In this combination, the disconnected graphs vanish exactly. Hence, we have to consider only one Wick contraction, which is schematically drawn in figure \ref{fig:nucleon-three-pt}. It is called $C_{3\mathrm{pt}}^{\Gamma,\mathrm{conn}}$ in the following.

The connected Wick contraction is evaluated using smeared quark sources at random spatial position $\vec{z}$. The propagator obtained by an inversion of this source is denoted by $M_z^{\Phi,\vec{P}}$. The superscript $\Phi,\vec{P}$ indicates that the source has been treated with momentum smearing \cite{Bali:2016lva} in combination with HYP smeared gauge links \cite{Hasenfratz:2001hp}.
Momentum smearing is crucial to realize large hadron momenta on the lattice, as
it increases the overlap with the proton ground state dramatically. The propagator connecting the proton sink and the insertion operator is evaluated by the sequential source method \cite{Maiani:1987by}. The proton sink and the sequential source, which is placed at time slice $t$, are again improved by momentum smearing. The corresponding sequential propagator is denoted by $X^{\Phi ,\vec{P}}_{t,3\mathrm{pt}}$. Notice that it should be Hermitian conjugated and followed by multiplication with $\gamma_5$, since an inversion on the sequential source returns a backward propagator. In total, we can write the connected contraction as:
\begin{align}
\label{eq:3pt-connected}
C_{3\mathrm{pt}}^{\Gamma,\mathrm{conn}}(\vec{P},S,\mathcal{C},t,\tau) = \left\langle e^{i\vec{P}\vec{z}} \sum_{\vec{x}} \left[ X^{\dagger,\Phi ,\vec{P}}_{t,3\mathrm{pt}} (\vec{x},\tau) \gamma_5 \Gamma \mathcal{U}^{\mathrm{latt}}[\mathcal{C}] M_z^{\Phi,\vec{P}} (\vec{x},\tau) \right] \right\rangle \,,
\end{align}
where $\mathcal{U}^{\mathrm{latt}}[\mathcal{C}]$ denotes the staple-shaped gauge link, which is constructed using the lattice gauge links \eqref{eq:lattice-gauge-links-straight} or \eqref{eq:lattice-gauge-links-step}, respectively. Notice that we are able to calculate $C_{3\mathrm{pt}}^{\Gamma,\mathrm{conn}}$ for all insertion time slices $\tau$ by performing only one sequential inversion, whereas the sink time slice $t$ is fixed by the location of the sequential source.

\subsection{Simulation setup}

The two-point function, as well as the connected three-point graph $C_{3\mathrm{pt}}^{\Gamma,\mathrm{conn}}$ are evaluated on $2000$ configurations of the H101 CLS ensemble \cite{Bruno:2014jqa}. It employs the tree-level improved L\"uscher-Weisz gauge action and includes $N_f = 2+1$ Sheikholeslami-Wohlert fermions. The extension is $32^3\times 96$ with lattice spacing $a= 0.0854$fm. Additional ensemble parameters are given in table~\ref{tab:cls-parameter-overview}.

\begin{table}[tbp]
  \centering
  \begin{tabular}{|l|l|l|l|l|l|l|l|l|l|l|}
    \hline
    name & $\beta$ & $L^3 \times T$ & $a[\mathrm{fm}]$ & $\kappa_l = \kappa_s$ & $m_{\pi}$ & $m_K$ & $m_{\pi}L$ & $L[\mathrm{fm}]$ & conf\\
    \hline
    H101 & $3.4$ & $32^3 \times 96$ & $0.0854$ & $0.13675962$ & $422$ & $422$ & $5.8$ & $2.7$ & $2016$ \\
    \hline
  \end{tabular}
  \caption{\label{tab:cls-parameter-overview} Parameters of the H101 CLS ensemble used in the present study.}
\end{table}

For each configuration, we perform calculations for four different quark sources located at random spatial position with the source time slice $t_{\mathrm{src}} = 10a$, and the source-sink separation $t_{\text{snk}}-t_{\text{src}}=11a$. The quantity that enters the l.h.s.\ of \eqref{eq:def-lattice-ratio} is obtained by a fit to a constant regarding the insertion time $\tau$. The fit includes an interval of insertion time slices assuming that excited states are sufficiently suppressed. In the present work we consider $\tau \in [4a,7a]$.

In our simulation the proton momentum $\vec P = (P_1, 0, 0)$ has values $P_1 \in \{0, 1, 2, 3\}\frac{2\pi}{aL}$. The first case $P_1 = 0$ is used to extract the nucleon mass and cross-check the dispersion relation. The other cases provide us with three values for $P^+=\{1.25,1.74,2.27\}$\,GeV, useful for the extraction of the CS kernel. The proton spin is oriented along the $z$-direction, i.e. we insert
\begin{align}
\label{eq:lattice-spin-proj}
\Gamma_S = \frac{1}{2}\left(1+\gamma^4\right)\left(1-i\gamma^2\gamma^1\right)
\end{align}
in \eqref{eq:3pt-def} and \eqref{eq:def-lattice-ratio}. Here, the first term is used to project the proton at rest onto the positive parity state. Since the proton momentum points in $x$-direction, the proton spin lies in the transverse plane, and thus $\lambda=0$. The vector $v$ is taken to be $v^\mu=(\pm 1,0,0,0)$, such that $(v\cdot P)=\pm P_1$. The vector $b$ is $b^\mu=(0,b_y,b_z,0)$. In the case $b_y\neq 0$ and $b_z\neq 0$, the transverse link has step-like form (\ref{eq:lattice-gauge-links-step}).

In order to reduce autocorrelation, we apply the binning method to the measured $C_{2\mathrm{pt}}$ and $C_{3\mathrm{pt}}^{\Gamma,\mathrm{conn}}$, with bin size $20$. From the resulting binned samples, we create $100$ jackknife samples, which are used for the estimation of the error propagation in all derived quantities, such as the ratio \eqref{eq:def-lattice-ratio}.

\subsection{Extraction of lattice correlators $W$}

To extract physical matrix elements, we parametrize the correlators $W^{[\Gamma]}$ using the most general parameterizations and fit the invariant structure functions. This approach has been developed in refs. \cite{Musch:2010ka,Musch:2011er}, and successfully applied to the computation of moments of various TMDs in the references \cite{Musch:2011er,Engelhardt:2015xja,Engelhardt:2017miy,Yoon:2017qzo,Engelhardt:2020qtg}. Note that, although this method can also be used to scan the $\ell$-dependence of the correlators and extract the $x$-dependence of TMDs, the application to first moments ($\ell=0$) is particularly simple and robust, and suited to the current task -- the extraction of the CS kernel.

The detailed description of this procedure with explicit expressions is given in ref. \cite{Musch:2010ka}. Here, for illustration purposes, we present only the parametrization for the vector matrix element. It reads
\begin{eqnarray}
\label{eq:parametrization_vector_channel}
\widetilde W^{[\gamma^\mu]}(b;\ell = 0, L;v,P,S) &=&
P^{\mu} \tilde{a}_2 + M^2 (Lv^{\mu}) \tilde{b}_1
- iM \epsilon^{\mu \nu \alpha \beta} P_{\nu} b_{\alpha} S_{\beta} \tilde{a}_{12}
\\\nn &&- i M^3 \epsilon^{\mu \nu \alpha \beta} b_{\nu} (Lv_{\alpha}) S_{\beta} \tilde{b}_8
- i M^2 b^{\mu} \tilde{a}_3
\\\nn &&
 + M \epsilon^{\mu \nu \alpha \beta} P_{\nu} (Lv_{\alpha}) S_{\beta} \tilde{b}_7 - M^3 (b \cdot S) \epsilon^{\mu \nu \alpha \beta} P_{\nu} b_{\alpha} (Lv_{\beta}) \tilde{b}_9
\\&&\nn
- i M^3 ((Lv) \cdot S) \epsilon^{\mu \nu \alpha \beta} P_{\nu} b_{\alpha} (Lv_{\beta}) \tilde{b}_{10},
\end{eqnarray}
where for our calculation at $\ell = 0$, $\tilde a_i$ and $\tilde b_i$ are functions of the Lorentz-invariant combinations $(b^2,L(v\cdot P), (Lv)^2)$ (since $(v\cdot b)=(b\cdot P)=0$). Using the set of matrix elements $\widetilde W^{[\gamma^\mu]}$ evaluated at different $\mu$, $L$ and $b^\nu$ we fit the invariant functions $\tilde a_i$ and $\tilde b_i$ with a least-square fit.

In the following step, the fitted functions $\tilde a_i$ and $\tilde b_i$ are grouped into the leading twist TMD combinations (\ref{def:leading-twist}). Using that the spin-vector is an independent vector we identify the combinations of $\tilde a_i$ and $\tilde b_i$ that refer to particular TMD distributions (\ref{def:vector},\ref{def:axial},\ref{def:tensor}). In the vector case, the terms proportional to the unpolarized ($f_1$) and Sivers ($f_{1T}^\perp$) TMDs are
\begin{eqnarray}
W^{[f_1]}(b^2,L,P^+)=P^+ \(\tilde a_2(b^2)+M^2 \frac{Lv^+}{P^+}\tilde b_1(b^2)\),
\\
W^{[f_{1T}^\perp]}(b^2,L,P^+)=P^+ \(\tilde a_{12}(b^2)-M^2 \frac{Lv^+}{P^+}\tilde b_8(b^2)\).
\end{eqnarray}
Here, we introduce the notation $W^{[F]}$, which indicates the component of the matrix element $W^{[\Gamma]}$ with the leading contribution proportional to the TMD $F$. In other words, it is a qTMD (at $\ell=0$) with the quantum numbers corresponding to $F$.

The axial and tensor qTMDs defined in (\ref{def:axial}) and (\ref{def:tensor}) are computed analogously. In our calculation we have access to the 6 leading-twist TMDs that remain present in (\ref{def:vector},\ref{def:axial},\ref{def:tensor}) after setting $\lambda = 0$ as implied by our choice of the proton spin. These are the unpolarized TMD $f_1$, the Sivers function $f_{1T}^\perp$, worm-gear-T function $g_{1T}$, transversity TMD $h_1$, Boer-Mulders function $h_1^\perp$ and pretzelocity $h_{1T}^\perp$. The evaluation of the functions $g_{1L}$ and $h_{1L}^\perp$ is also possible but requires simulation with a different spin orientation.

\section{The CS kernel from lattice data}
\label{sec:CS}

In this section, we present the CS kernel extraction from the $W^{[F]}$ determined by the lattice simulations. We also estimate the systematic uncertainty and compare the results of the extraction with previous extractions.

\begin{figure}[t]
\centering
\includegraphics[width=.95\textwidth]{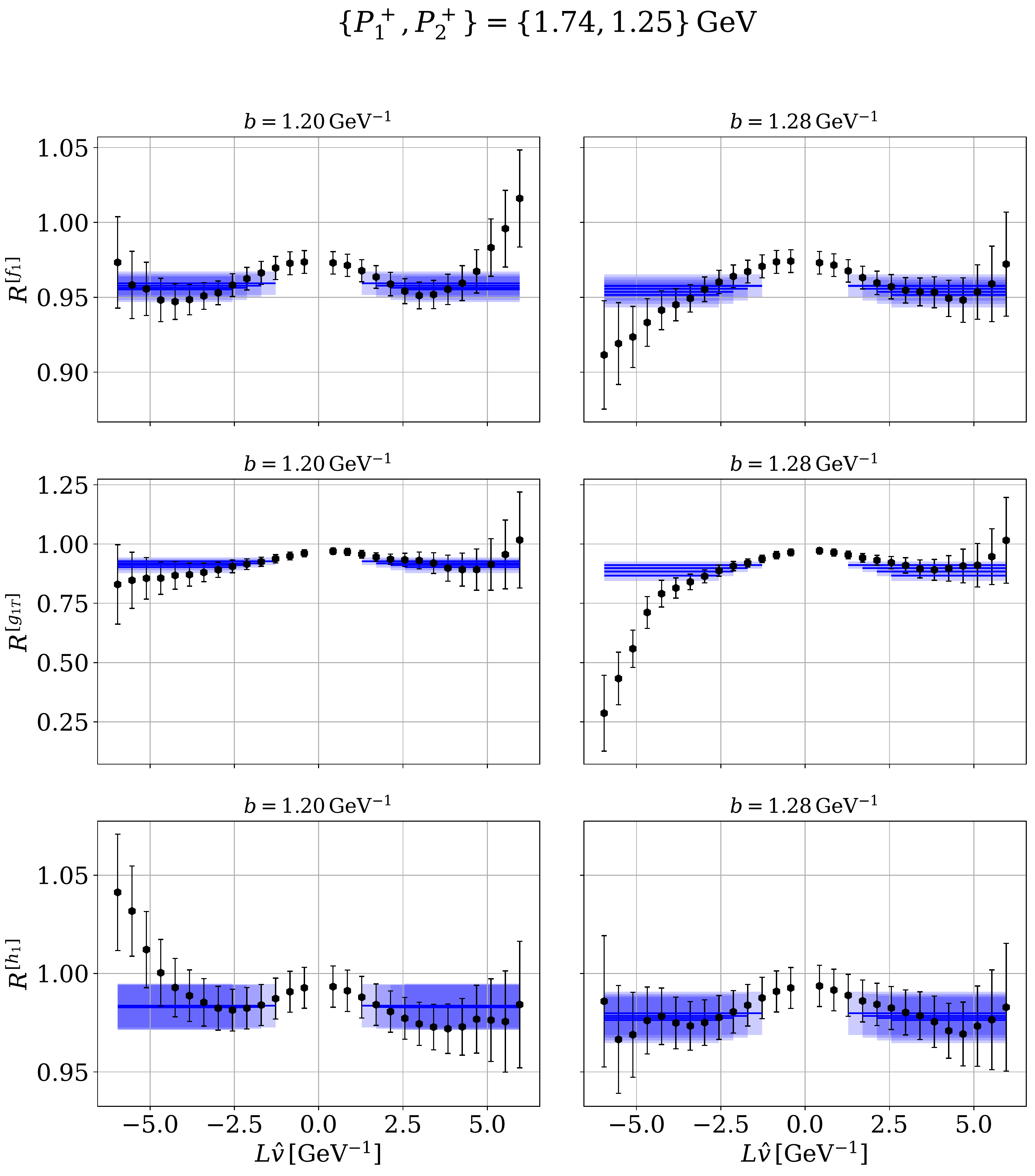}
\caption{\label{fig:plot_R_lattice} Lattice results for the ratios between different momenta including the plateau fit for different fit ranges including the jackknife error of the fit. The shown values of $b$ are used to determine $\mathbf{M}^{[\Gamma]}$ from these ratios.}
\end{figure}

\subsection{Determination of the CS kernel}
\label{sec:dermination-of-CS}

As an outcome of the lattice simulation, described in the previous section, we have the functions $W^{[F]}$ evaluated at different values of $L$, $(v\cdot P)$, $P^+$ and $|\vec b|$. From these functions we construct the ratios (\ref{th:ratio0})
\begin{equation}\label{th:ratio-latt}
R_{\textnormal{latt}}^{[F]}(P_1, P_2; b; L; \hat v) = \frac{W^{[F]}(P_1; b; L; v)}{W^{[F]}(P_2; b; L; v)},
\end{equation}
where $\hat v=\sign(v\cdot P)$. It is important that many lattice-related artifacts, such as lattice renormalization factors, cancel in this ratio. This cancellation is not exact since there is an operator mixing (such as discussed in ref. \cite{Shanahan:2019zcq}) that possibly depends on momentum. This point requires further investigation, which will be performed in the future. For the moment, we ignore these effects, expecting them to be small in comparison to other distortions, such as lattice artifacts and power corrections to the factorization theorem.

\begin{figure}[t]
\centering 
\includegraphics[width=.95\textwidth]{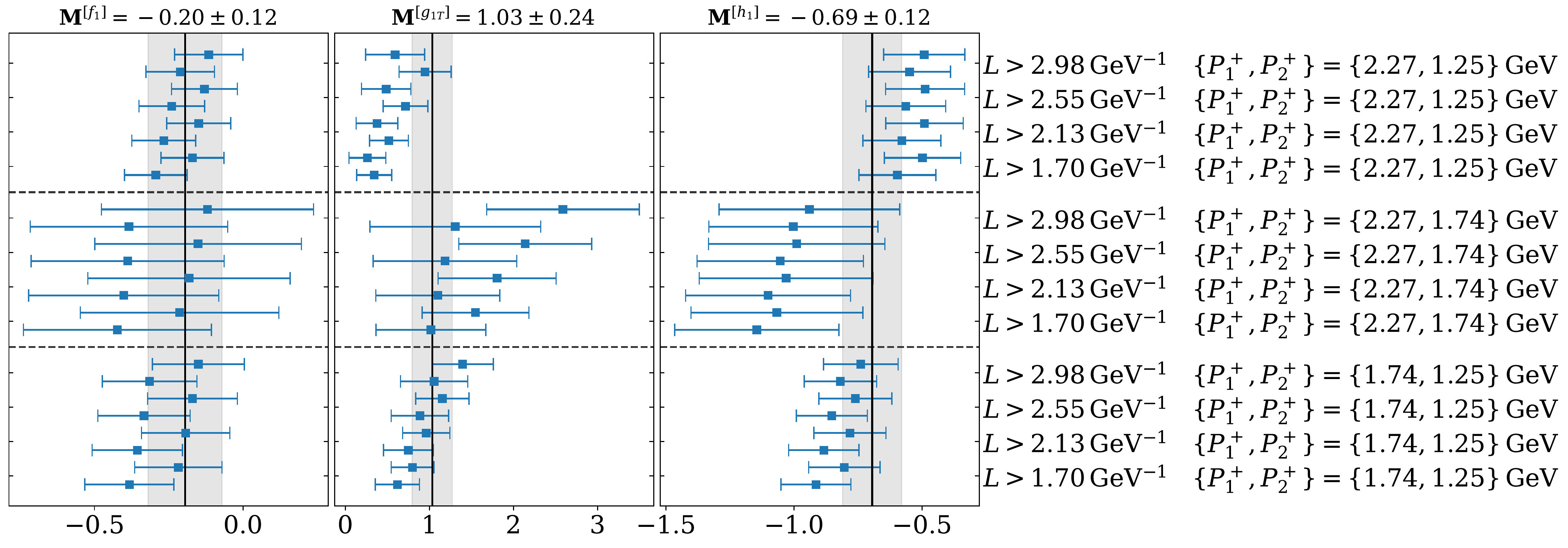}
\caption{\label{fig:plot_m} Fit results for $\mathbf{M}^{[F]}$ for different combinations of momenta $\{P_1, P_2\}$ and cutoffs of the staple length $L$. Note that the results for the two largest lattice momenta $\{P_1^+,P_2^+\} = \{ 2.27, 1.74\}\,\mathrm{GeV}$ have the largest statistical error, which can be explained with a more severe violation of Lorentz invariance and the dispersion relation at larger momenta.}
\end{figure}

We analyze the ratios (\ref{th:ratio-latt}) using the theoretical scheme presented in sec.\ref{sec:theory}, and extract the CS kernel. The analysis has three principal steps, which we describe below.

\textit{(i) Extrapolation $L\to\infty$.} The large-$L$ limit is an essential requirement to reach the TMD regime (\ref{th:fac-limit}). To extract the large-$L$ value of $R^{[F]}$ we use a constant fit (for $L>2.98$\,GeV$^{-1}$) independently for each combination of $b$, $P_1^+/P_2^+$ and $F$. Example profiles of $R^{[F]}$ in $L$ and extracted plateau values are shown in figure \ref{fig:plot_R_lattice}. In the plateau fit we assume that ratios at the values for $\hat v = \pm 1$ are the same (which is the consequence of parity conservation) and perform a combined fit for the left and right branches.

In many cases, the profile in $L$ does not have a well-defined transition to an asymptote. Such a situation is especially typical for large values of $b$ ($b>2.5$\,GeV$^{-1}$). It indicates that the values of $L$ used in our analysis are not large enough, and corrections $\sim b/L$ are significant. This part of the analysis provides the largest source of uncorrelated uncertainties. In the following plots, we specially shade the area $b>2.5$\,GeV$^{-1}$ to indicate that this part of the extraction is not under control.

We faced problems with the extrapolation of $R^{[F]}$ for the Sivers $f_{1T}^\perp$, Boer-Mulders $h_{1}^\perp$ and pretzelocity $h_{1T}^\perp$ cases. Particularly, the ratios $R^{[f_{1T}^\perp]}$ and $R^{[h_1^\perp]}$ show anomalous growth at large $b$. The origin of this problem is not entirely clear. We associate it with the incomplete cancellation of lattice artifacts. Alternatively, such abnormal behavior can be described by a significant violation of the $\mathbf{M}=\const$ assumption, see also fig.~\ref{fig:2}\,(right). The ratio $R^{[h_{1T}^\perp]}$ has a very small signal-to-noise value. For these reasons, we exclude these three cases\footnote{Notice that these cases are the only TMDs that do not match twist-two distributions at small-$b$, but twist-three distributions \cite{Moos:2020wvd}. The general smallness of higher twist corrections explains the large noise issue. Additionally, the Sivers and Boer-Mulders functions are P-odd and have leading contributions of the Qiu-Sterman type \cite{Qiu:1991pp}, with a zero-momentum gluon. In these cases, one could expect that power corrections in $L^{-1}$ are larger.} from the following consideration, which left us with three TMDs.

\textit{(ii) Determination of $\mathbf{M}$.} The phenomenological formula for the extraction of the CS kernel contains the unknown function $\mathbf{M}$, which corrects the ratio for the unobserved $x$-dependence. In our approach we assume $\mathbf{M}(b)=\const$, and extract it from $R^{[F]}_{\text{latt}}$. The extraction is performed at values of $b$ which, on the one hand, are small enough to be described by perturbative QCD, and on the other hand, are large enough to satisfy (\ref{th:fac-limit}). We choose $b = 1.20$\,GeV$^{-1}$ or $1.28$\,GeV$^{-1}$ (for configurations which contain these points), and fit the value of $\mathbf{M}$ comparing $R^{[F]}$ to equation (\ref{th:ratio-cs-kernel}) with the perturbative $\mathbf{r}$ given in (\ref{th:r-pert}). The perturbative CS kernel used in this comparison is taken at N$^3$LO \cite{Vladimirov:2016dll}. For better control of the extraction uncertainty we perform an additional fit-range variation during the $L\to\infty$ extrapolation, and consider the cases $L>\{1.70,2.13,2.55,2.98\}$\,GeV$^{-1}$. The extracted values of $\mathbf{M}$ are presented in fig.~\ref{fig:plot_m}.

The determination of $\mathbf{M}$ is the central element of our analysis. According to our hypothesis, the values of $\mathbf{M}$ must be constant and depend only on the type of TMD. The deviation from a single value for different extractions provides an additional source of uncertainty in our analysis. To account for this uncertainty, we use the average value of $\mathbf{M}$ computed for each set of TMDs, as is shown in fig.~\ref{fig:plot_m}. Generally, we found a good agreement between different momentum configurations and different fitting ranges. Partially, the difference between different momentum configurations is explained by the evolution effects for $\mathbf{M}$, which are subleading in our analysis.

\textit{(iii) Extraction of $K$}. We combine the extrapolated $R^{[F]}_{\text{latt}}(L\to\infty)$ with the values of $\mathbf{M}$ and determine the CS kernel according to equation (\ref{th:cs-kernel-explicit}). This gives us the CS kernel at the factorization scale $\mu$ (\ref{th:mu}). For cross-comparison and comparison with other extractions we evolve CS kernels to $\mu=2\,\textnormal{GeV}$. The evolution equation (\ref{th:integrability}) is independent of $b$ and thus gives us a flat shift for the CS kernel value, $K(b,2\,\textnormal{GeV})=K(b,\mu)+\Delta(\mu)$. 
The values of the factorization scale $\mu$ and $\Delta$ are summarized in the following table,
\begin{center}
\begin{tabular}{|c||c|c|c|}
\hline
$\{P_1^+, P_2^+\}$ & \{1.74,1.25\}GeV & \{2.27,1.74\}GeV  & \{2.27,1.25\}GeV
\\\hline
$\mu$ & 2.09 GeV & 2.81 GeV & 2.38 GeV
\\\hline
$\Delta(\mu)$ & 0.013 & 0.053 & 0.098\\
\hline
\end{tabular}
\end{center}

In all steps the statistical propagation of uncertainties is performed with the resampling method. All theoretical predictions, such as the perturbative CS kernel, evolution factors, etc, are evaluated by \texttt{artemide} \cite{Scimemi:2017etj} at N$^3$LO with free parameters tuned as in the SV19 TMD-extraction \cite{Scimemi:2019cmh}.

\subsection{Estimation of systematic uncertainties}
\label{sec:uncertanties}

\begin{figure}[t]
\centering 
\includegraphics[width=.95\textwidth]{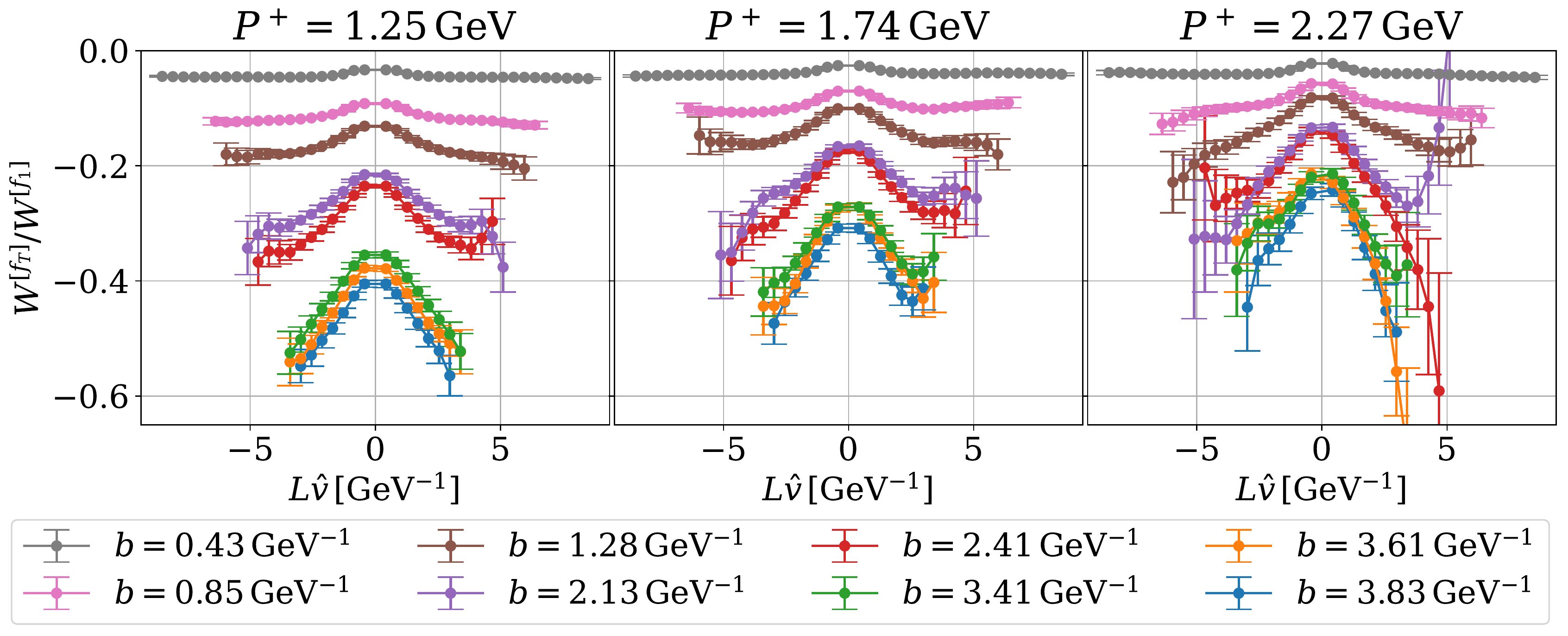}
\caption{\label{fig:plot_a3ratios_p1} Ratios $\frac{f_{T}(P^+; L\hat{v}; b)}{f_1(P^+; L\hat{v}, b)}$ for different values of $P^+$. The twist-three TMD $f_T$ parameterizes the $\sim b^\mu$ component of the vector current $g_T^{\mu\nu}\bar q\gamma_\nu q$.}\end{figure}

\begin{figure}[t]
\centering 
\includegraphics[width=.95\textwidth]{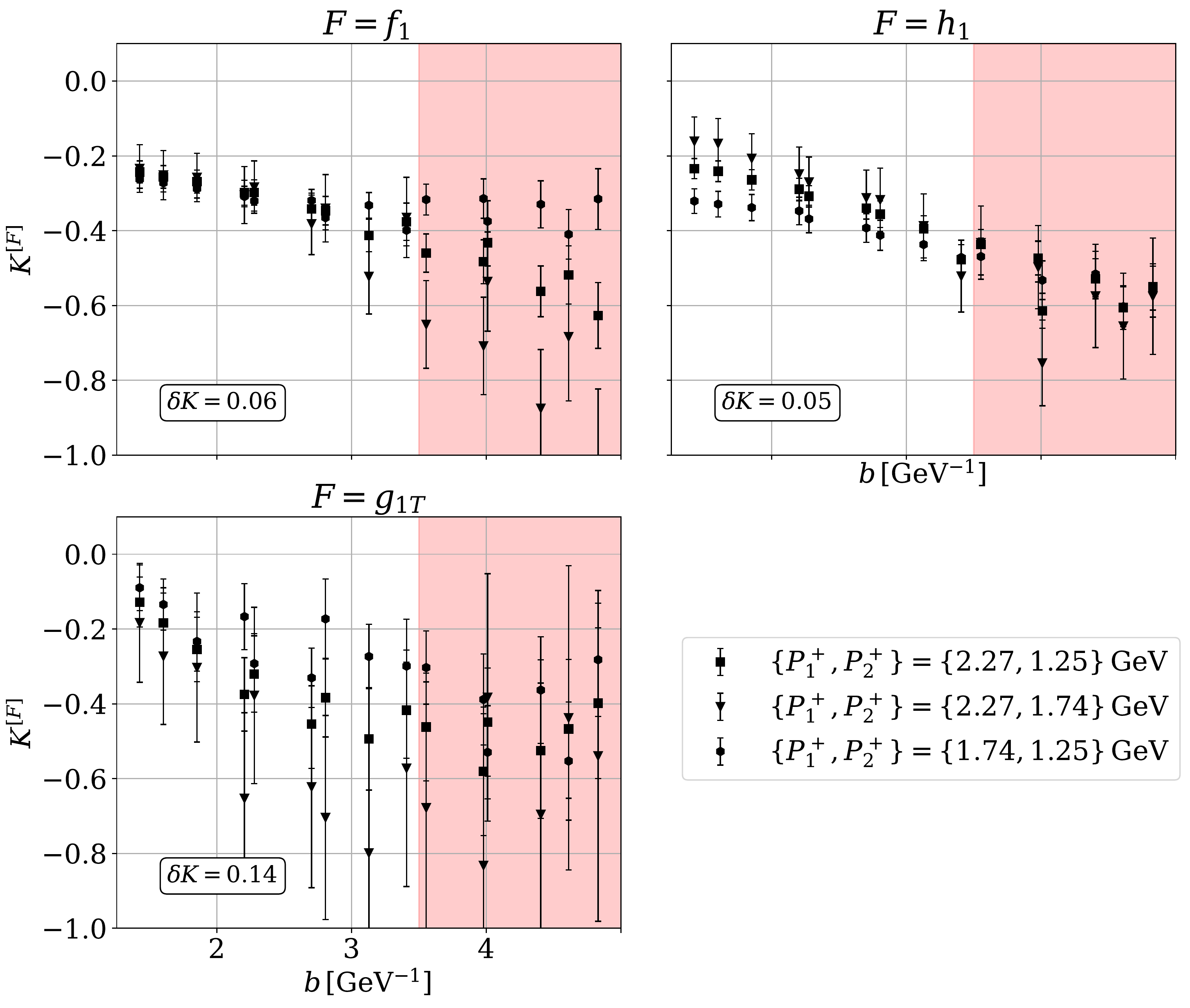}
\caption{\label{fig:h101_cs_kernel_teven_different_momenta} Values for the CS kernel $K(b,\mu=2\text{GeV})$ extracted in different TMD channels using equation~\eqref{th:cs-kernel-explicit}. The red shaded area indicates the region where large power corrections have to be expected (see discussion in sec.\ref{sec:uncertanties}). The values $\delta K$ represent estimates of the correlated systematic uncertainty due to model assumptions.}
\end{figure}

\begin{figure}[t]
\centering 
\includegraphics[width=.95\textwidth]{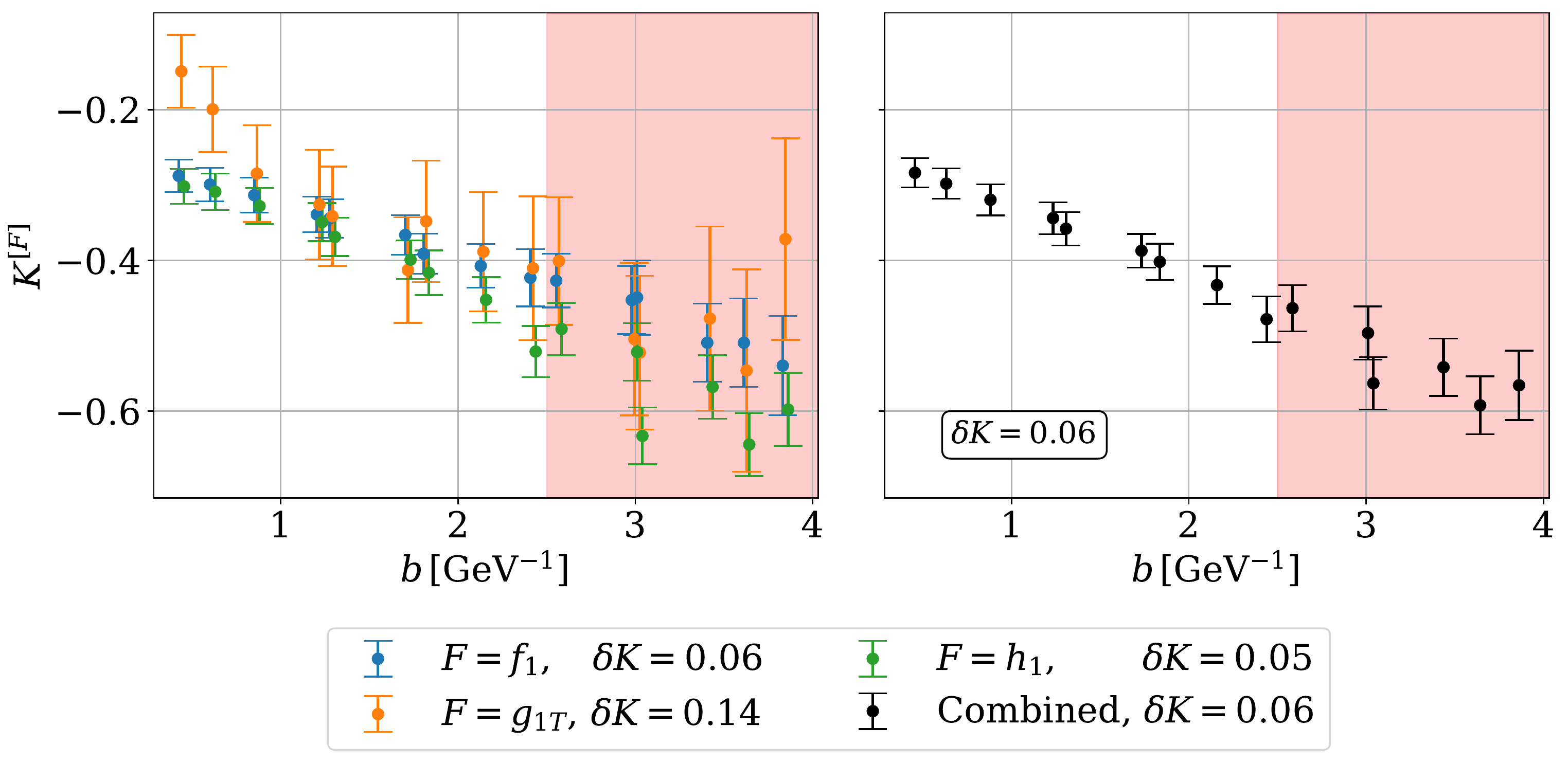}
\caption{\label{fig:K-momentum-avaraged} Values for the CS kernel $K(b,\mu=2\,\text{GeV})$ combined from the data in fig.\ref{fig:h101_cs_kernel_teven_different_momenta}.}
\end{figure}

The results of extraction are given in figs. \ref{fig:h101_cs_kernel_teven_different_momenta} and \ref{fig:K-momentum-avaraged} and discussed in the following section. The error-bars on these figures represent the statistical uncertainty only. The accurate estimation of systematic uncertainty is impossible at the current stage of research. However, let us at least identify the main sources and give an estimate of possible effects.
\begin{itemize}
\item \textit{Lattice artifacts.} Presumably, this is the main source of uncertainty and the least controlled. Mainly, the lattice artifacts arise due to differences between proton states evaluated at different momenta. This effect is visible in fig.~\ref{fig:plot_R_lattice} as a sudden change of asymptotics at large-$L$. In some cases ($f_{1T}^\perp$, $h_1^\perp$, $h_{1T}^\perp$) these lattice artifacts completely ruin the extraction. An improved understanding of these artifacts and their control is essential for further development.

\item \textit{Assumption $\mathbf{M}=\const$.} This assumption introduces a fully correlated systematic uncertainty, the size of which is estimated by equation (\ref{th:deltaK1}). The values of $\delta\mathbf{M}$ are given in fig.~\ref{fig:plot_m}, and the resulting values for $\delta K$ are indicated on each figure.

\item \textit{Small size of staple contour.} The insufficient size of the contour generates a systematic deformation of the CS kernel at large $b$. We expect that the values with $b>2.5\,\textnormal{GeV}^{-1}$ are systematically higher (lower for $|K|$). This effect occurs because many configurations do not reach the plateau for available values of $L$, and thus corresponding values are systematically underestimated. Thus, in fig. \ref{fig:h101_cs_kernel_teven_different_momenta} and \ref{fig:K-momentum-avaraged} this area is shaded in red. The same effect but in a smaller amount takes place also at smaller values of $b$ in some configurations. This is apparent in fig.~\ref{fig:plot_m}, where some series of points are moving to the left or right with an increase of $L$, e.g. for $g_{1T}$ at $P_1^+=2.27\,\mathrm{GeV}$.

\item \textit{Power corrections in $P$ to factorization theorem.} The factorization theorem is valid at large $P$, whereas the available values are not that large. The amount of violation of the factorization theorem can be estimated by the method suggested in ref. \cite{Vladimirov:2020ofp}, which uses the fact that in the factorization limit, only the ``good'' components of the quark fields survive (\ref{def:leading-twist}). In turn, it implies that certain tensor components, cf. (\ref{th:gamma-prime}), of the quark-bilinear are larger than others. Therefore, the ratio of a ``bad'' component to a ``good'' component provides an estimation for the size of power corrections to the factorization theorem. We performed such a comparison and found that these corrections amount to $\sim 40\%$ at $P^+=1.25\,\textnormal{GeV}$, and $\sim 30\%$ at $P^+=2.27\,\textnormal{GeV}$ for $b\simeq 2.5\,\textnormal{GeV}^{-1}$. In fig.~\ref{fig:plot_a3ratios_p1} we demonstrate the case $W[f_T]/W[f_1]$, where $f_T$ is the twist-three TMD that parametrizes the $\sim b^\mu$ component of the  vector current ($g_T^{\mu\nu}\langle \bar q\gamma_\nu q\rangle \sim b^\mu M f_T$, ignoring common structures). The tested ratios exhibit the expected general behavior, i.e., they decrease with the increase of $P^+$ or decrease of $b$. Fortunately, these corrections do not significantly impact our extraction because they mostly cancel in the ratio $R^{[F]}$ and are partially accounted for by the fit of $\mathbf{M}$. However, these corrections will present significant problems for determining absolute values for TMDs from the lattice.

\item \textit{Power corrections in $b^{-1}$ to factorization theorem.} At small values of $b$ the factorization theorem is violated by $1/|b|P^+$ corrections. Therefore, the range of small-b is not reliable. The effect of these corrections is apparent for $b\lesssim 0.8\,\textnormal{GeV}^{-1}$, where the CS kernel is perturbative.
\end{itemize}
There are also numerous smaller sources of systematic uncertainties, which are not discussed here. In general, we believe that our extraction, supplemented with the statistical and the given correlated uncertainties, provides a reliable estimation for the CS kernel in the range $0.8 \lesssim b \lesssim 2.5\,\textnormal{GeV}^{-1}$.

\subsection{Discussion}
\label{sec:discussion}

The extracted values of the CS kernel from three different TMDs, and three different momentum combinations are shown in fig.~\ref{fig:h101_cs_kernel_teven_different_momenta}. Clearly, the present uncertainties are dominated by statistical noise. Nonetheless, we observe a good agreement between different channels, which confirms the CS kernel's universality. The noisiest channel is the worm-gear TMD $g_{1T}$. In this case, the main source of disagreement comes from the spread of $\mathbf{M}$ (see fig.~\ref{fig:plot_m}). The uncertainty in the determination of $\mathbf{M}$ gives rise to the correlated uncertainty, which is denoted by $\delta K$ and indicated for each case.

In fig.~\ref{fig:K-momentum-avaraged} (left panel), we present the values of the CS kernel from the combined fit of different momentum configurations for each TMD. Effectively, the averaging over channels triples the statistics. We observe that the extractions from $f_1$ and $h_1$ are in excellent agreement with each other, whereas the curves we obtain from $g_{1T}$ have a rising tendency at large $b$. This effect mainly appears due to the small sizes of the contour. The right panel of fig.~\ref{fig:K-momentum-avaraged} shows the values for the CS kernel combined from all channels and represents the main result of our work. The combined value of $\delta K$ is computed according to ref.~\cite{Schmelling:1994pz}.

\begin{figure}[tbp]
\centering 
\includegraphics[width=.95\textwidth]{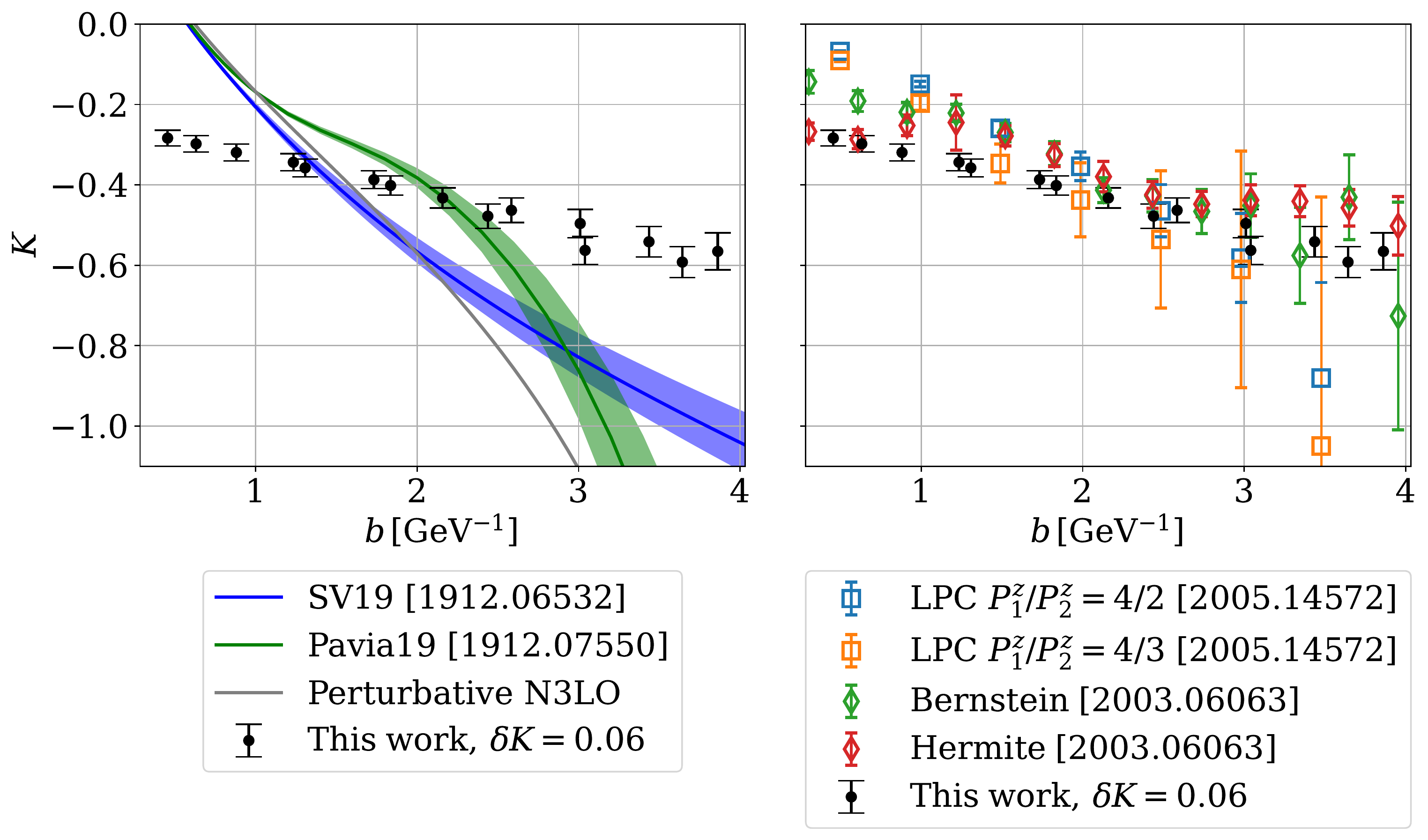}
\caption{\label{fig:plot_cs_kernel_other_extractions} Comparison of the CS kernel extracted in this work to phenomenological extractions (left) and lattice computations (right). The comparison is made at $\mu=2$\,GeV.}
\end{figure}

We compare the obtained CS kernel with other results in fig.~\ref{fig:plot_cs_kernel_other_extractions}. The left panel shows the comparison with phenomenological extractions \cite{Scimemi:2019cmh,Bacchetta:2019sam} and the purely perturbative CS kernel resummed at N$^3$LO \cite{Vladimirov:2016dll,Echevarria:2012pw}. We observe some similarity between extractions in the range $1<b<2.5$\,GeV$^{-1}$ (which is the most reliable part of our analysis). However, for $b>2.5$\,GeV$^{-1}$ the discrepancy is essential, as well as the difference between the two phenomenological curves. Probably this indicates that these curves are primarily determined by the chosen analytic form of the parametrizations for such large $b$. The right panel of fig.~\ref{fig:plot_cs_kernel_other_extractions} shows the comparison with other lattice computations, which were made in ref.~\cite{Shanahan:2020zxr} (in the quenched approximation) and in ref.~\cite{Zhang:2020dbb} (where the CS kernel is extracted from the qTMD soft factor \cite{Ji:2019sxk}). We observe a nice agreement.

Importantly, all lattice simulations demonstrate a weak variation of the CS kernel at large-$b$. This observation contradicts the popular assumption $K(b)\sim b^2$ for large $b$, see e.g. refs.~\cite{Bacchetta:2019sam,Landry:2002ix,Su:2014wpa}, and supports models with linear or constant asymptotics, such as in ref.~\cite{Collins:2014jpa,Scimemi:2019cmh,Vladimirov:2020umg}.

\section{Conclusion}
\label{sec:conclusion}

In this work, we present a lattice computation of the Collins-Soper (CS) kernel from the first moments of qTMD distributions. The analysis is performed with dynamical fermions and staple-shaped Wilson lines for the CLS ensemble H101 with lattice spacing $a = 0.0854$\,fm and unphysical quark masses. The results are shown in figures \ref{fig:h101_cs_kernel_teven_different_momenta} and \ref{fig:K-momentum-avaraged}. For the first time the CS kernel is determined from three different TMDs ($f_1$, $g_{1T}$, $h_1$). The results agree with each other, which confirms the universality of the approach.

The method of extraction used in this work was suggested in ref.~\cite{Vladimirov:2020ofp} and is based on the NLO factorization formula for qTMDs \cite{Ebert:2018gzl,Ebert:2019okf,Ji:2019ewn,Vladimirov:2020ofp}. In contrast to other methods \cite{Ebert:2018gzl,Ebert:2019okf,Ji:2019ewn}, it does not require evaluating the $x$-dependence of qTMDs but uses only $x$-moments. The information on the $x$-dependence is contained in a certain integral combination of TMDs, called $\mathbf{M}$ (\ref{th:M-def}). The central assumption of the approach is that $\mathbf{M}$ is independent of $b$, which we confirm (within statistical uncertainties) from our lattice data. The ambiguity in the determination of $\mathbf{M}$ is contained in the fully correlated uncertainty $\delta K$. The given values for $\delta K$ may be exceeded by other (not estimated) systematic uncertainties. Overall, our results demonstrate that the method suggested in ref.~\cite{Vladimirov:2020ofp} is suitable for determining the CS kernel.

Our extraction of $K$ is supplemented with a discussion of sources of systematic uncertainties. We identify the pure lattice artifacts as the largest and the least controlled source of uncertainty. Excessive lattice artifacts force us to exclude the functions $f_{1T}^\perp$, $h_1^\perp$ and $h_{1T}^\perp$ from our analysis. Other sources, such as power corrections to qTMD factorization and model assumptions, produce smaller and better-quantified uncertainties. We identify as the reliable range for the extracted CS kernel the interval $0.8\,\text{GeV}^{-1}\lesssim b \lesssim 2.5$\,GeV$^{-1}$. We also estimate the size of power suppressed corrections to TMD factorization and find them to amount to $\sim 30-40\%$ at our energies. These corrections do not strongly influence the present extraction of the CS kernel but will play a vital role in extracting the $x$-dependent TMDs.

In fig.~\ref{fig:plot_cs_kernel_other_extractions} we demonstrate that our results are in agreement with previous lattice calculations \cite{Shanahan:2020zxr,Zhang:2020dbb}. In our case, the statistical uncertainty is smaller and comparable to the uncertainty of phenomenological extractions due to a larger number of channels for determining the CS kernel. For that reason, and because we can better estimate systematic uncertainties, we regard our result as a significant step forward in the lattice determination of the CS kernel. The comparison with the phenomenological extraction exhibits discrepancies, which are mainly due to systematic sources. Therefore, future studies should concentrate on the elimination of systematic contaminations.

\acknowledgments
{We acknowledge PRACE for awarding us access to SuperMUC-NG at GCS@LRZ, Germany.
The authors also gratefully acknowledge the Gauss Centre for Supercomputing e.V. (www.gauss-centre.eu) directly for manifold help and funding this project by providing additional computing time in the start-up phase SuperMUC-NG at Leibniz Supercomputing Centre (www.lrz.de). The authors thank the Rechenzentrum of Regensburg for providing the Athene Cluster for supplementary computations. The authors have greatly profited from discussions with Piotr Korcyl, Rajan Gupta, Bernhard Musch, Ignazio Scimemi, and Jeremy Green. This project was funded in part by DFG, SFB/TRR-55 ``Hadron Physics from Lattice QCD''. Michael Engelhardt is supported by the U.S. Department of Energy, Office of Science, Office of Nuclear Physics through grant DE-FG02-96ER40965 and through the TMD Topical Collaboration. The CLS collaboration is acknowledged for generating the $n_f=2+1$ ensembles. We also thank the authors of the Chroma software system~\cite{EDWARDS2005832} which we used to measure the correlators and the authors of Matplotlib~\cite{Hunter:2007} which was used for the visualization of our data.}

\bibliography{bibFILE}

\providecommand{\href}[2]{#2}\begingroup\raggedright\begin{thebibliography}{10}

\bibitem{Gautheron:2010wva}
{\scshape COMPASS} collaboration, F.~Gautheron et~al., \emph{{COMPASS-II
  Proposal}}, .

\bibitem{Aschenauer:2015eha}
E.-C. Aschenauer et~al., \emph{{The RHIC SPIN Program: Achievements and Future
  Opportunities}},  \href{https://arxiv.org/abs/1501.01220}{{\ttfamily
  1501.01220}}.

\bibitem{Dudek:2012vr}
J.~Dudek et~al., \emph{{Physics Opportunities with the 12 GeV Upgrade at
  Jefferson Lab}},
  \href{http://dx.doi.org/10.1140/epja/i2012-12187-1}{\emph{Eur. Phys. J. A}
  {\bfseries 48} (2012) 187},
  [\href{https://arxiv.org/abs/1208.1244}{{\ttfamily 1208.1244}}].

\bibitem{AbdulKhalek:2021gbh}
R.~Abdul~Khalek et~al., \emph{{Science Requirements and Detector Concepts for
  the Electron-Ion Collider: EIC Yellow Report}},
  \href{https://arxiv.org/abs/2103.05419}{{\ttfamily 2103.05419}}.

\bibitem{Braun:2007wv}
V.~Braun and D.~M\"uller, \emph{{Exclusive processes in position space and the
  pion distribution amplitude}},
  \href{http://dx.doi.org/10.1140/epjc/s10052-008-0608-4}{\emph{Eur. Phys. J.
  C} {\bfseries 55} (2008) 349--361},
  [\href{https://arxiv.org/abs/0709.1348}{{\ttfamily 0709.1348}}].

\bibitem{Ji:2013dva}
X.~Ji, \emph{{Parton Physics on a Euclidean Lattice}},
  \href{http://dx.doi.org/10.1103/PhysRevLett.110.262002}{\emph{Phys. Rev.
  Lett.} {\bfseries 110} (2013) 262002},
  [\href{https://arxiv.org/abs/1305.1539}{{\ttfamily 1305.1539}}].

\bibitem{Ma:2014jla}
Y.-Q. Ma and J.-W. Qiu, \emph{{Extracting Parton Distribution Functions from
  Lattice QCD Calculations}},
  \href{http://dx.doi.org/10.1103/PhysRevD.98.074021}{\emph{Phys. Rev. D}
  {\bfseries 98} (2018) 074021},
  [\href{https://arxiv.org/abs/1404.6860}{{\ttfamily 1404.6860}}].

\bibitem{Radyushkin:2017cyf}
A.~Radyushkin, \emph{{Quasi-parton distribution functions, momentum
  distributions, and pseudo-parton distribution functions}},
  \href{http://dx.doi.org/10.1103/PhysRevD.96.034025}{\emph{Phys. Rev. D}
  {\bfseries 96} (2017) 034025},
  [\href{https://arxiv.org/abs/1705.01488}{{\ttfamily 1705.01488}}].

\bibitem{Collins:1981va}
J.~C. Collins and D.~E. Soper, \emph{{Back-To-Back Jets: Fourier Transform from
  B to K-Transverse}},
  \href{http://dx.doi.org/10.1016/0550-3213(82)90453-9}{\emph{Nucl. Phys. B}
  {\bfseries 197} (1982) 446--476}.

\bibitem{Chiu:2011qc}
J.-y. Chiu, A.~Jain, D.~Neill and I.~Z. Rothstein, \emph{{The Rapidity
  Renormalization Group}},
  \href{http://dx.doi.org/10.1103/PhysRevLett.108.151601}{\emph{Phys. Rev.
  Lett.} {\bfseries 108} (2012) 151601},
  [\href{https://arxiv.org/abs/1104.0881}{{\ttfamily 1104.0881}}].

\bibitem{Vladimirov:2020umg}
A.~A. Vladimirov, \emph{{Self-contained definition of the Collins-Soper
  kernel}}, \href{http://dx.doi.org/10.1103/PhysRevLett.125.192002}{\emph{Phys.
  Rev. Lett.} {\bfseries 125} (2020) 192002},
  [\href{https://arxiv.org/abs/2003.02288}{{\ttfamily 2003.02288}}].

\bibitem{Scimemi:2019cmh}
I.~Scimemi and A.~Vladimirov, \emph{{Non-perturbative structure of
  semi-inclusive deep-inelastic and Drell-Yan scattering at small transverse
  momentum}}, \href{http://dx.doi.org/10.1007/JHEP06(2020)137}{\emph{JHEP}
  {\bfseries 06} (2020) 137},
  [\href{https://arxiv.org/abs/1912.06532}{{\ttfamily 1912.06532}}].

\bibitem{Bacchetta:2019sam}
A.~Bacchetta, V.~Bertone, C.~Bissolotti, G.~Bozzi, F.~Delcarro, F.~Piacenza
  et~al., \emph{{Transverse-momentum-dependent parton distributions up to
  N$^{3}$LL from Drell-Yan data}},
  \href{http://dx.doi.org/10.1007/JHEP07(2020)117}{\emph{JHEP} {\bfseries 07}
  (2020) 117}, [\href{https://arxiv.org/abs/1912.07550}{{\ttfamily
  1912.07550}}].

\bibitem{Bertone:2019nxa}
V.~Bertone, I.~Scimemi and A.~Vladimirov, \emph{{Extraction of unpolarized
  quark transverse momentum dependent parton distributions from
  Drell-Yan/Z-boson production}},
  \href{http://dx.doi.org/10.1007/JHEP06(2019)028}{\emph{JHEP} {\bfseries 06}
  (2019) 028}, [\href{https://arxiv.org/abs/1902.08474}{{\ttfamily
  1902.08474}}].

\bibitem{Ebert:2018gzl}
M.~A. Ebert, I.~W. Stewart and Y.~Zhao, \emph{{Determining the Nonperturbative
  Collins-Soper Kernel From Lattice QCD}},
  \href{http://dx.doi.org/10.1103/PhysRevD.99.034505}{\emph{Phys. Rev. D}
  {\bfseries 99} (2019) 034505},
  [\href{https://arxiv.org/abs/1811.00026}{{\ttfamily 1811.00026}}].

\bibitem{Ebert:2019okf}
M.~A. Ebert, I.~W. Stewart and Y.~Zhao, \emph{{Towards Quasi-Transverse
  Momentum Dependent PDFs Computable on the Lattice}},
  \href{http://dx.doi.org/10.1007/JHEP09(2019)037}{\emph{JHEP} {\bfseries 09}
  (2019) 037}, [\href{https://arxiv.org/abs/1901.03685}{{\ttfamily
  1901.03685}}].

\bibitem{Ji:2019ewn}
X.~Ji, Y.~Liu and Y.-S. Liu, \emph{{Transverse-Momentum-Dependent Parton
  Distribution Functions from Large-Momentum Effective Theory}},
  \href{http://dx.doi.org/10.1016/j.physletb.2020.135946}{\emph{Phys. Lett. B}
  {\bfseries 811} (2020) 135946},
  [\href{https://arxiv.org/abs/1911.03840}{{\ttfamily 1911.03840}}].

\bibitem{Vladimirov:2020ofp}
A.~A. Vladimirov and A.~Sch{\"a}fer, \emph{{Transverse momentum dependent
  factorization for lattice observables}},
  \href{http://dx.doi.org/10.1103/PhysRevD.101.074517}{\emph{Phys. Rev. D}
  {\bfseries 101} (2020) 074517},
  [\href{https://arxiv.org/abs/2002.07527}{{\ttfamily 2002.07527}}].

\bibitem{Lin:2020rut}
M.~Constantinou et~al., \emph{{Parton distributions and lattice QCD
  calculations: toward 3D structure}},
  \href{https://arxiv.org/abs/2006.08636}{{\ttfamily 2006.08636}}.

\bibitem{Ji:2020ect}
X.~Ji, Y.-S. Liu, Y.~Liu, J.-H. Zhang and Y.~Zhao, \emph{{Large-Momentum
  Effective Theory}},  \href{https://arxiv.org/abs/2004.03543}{{\ttfamily
  2004.03543}}.

\bibitem{Musch:2010ka}
B.~Musch, P.~H{\"a}gler, J.~Negele and A.~Sch{\"a}fer, \emph{{Exploring quark
  transverse momentum distributions with lattice QCD}},
  \href{http://dx.doi.org/10.1103/PhysRevD.83.094507}{\emph{Phys. Rev. D}
  {\bfseries 83} (2011) 094507},
  [\href{https://arxiv.org/abs/1011.1213}{{\ttfamily 1011.1213}}].

\bibitem{Musch:2011er}
B.~Musch, P.~H{\"a}gler, M.~Engelhardt, J.~Negele and A.~Sch{\"a}fer,
  \emph{{Sivers and Boer-Mulders observables from lattice QCD}},
  \href{http://dx.doi.org/10.1103/PhysRevD.85.094510}{\emph{Phys. Rev. D}
  {\bfseries 85} (2012) 094510},
  [\href{https://arxiv.org/abs/1111.4249}{{\ttfamily 1111.4249}}].

\bibitem{Engelhardt:2015xja}
M.~Engelhardt, P.~H\"agler, B.~Musch, J.~Negele and A.~Sch\"afer,
  \emph{{Lattice QCD study of the Boer-Mulders effect in a pion}},
  \href{http://dx.doi.org/10.1103/PhysRevD.93.054501}{\emph{Phys. Rev. D}
  {\bfseries 93} (2016) 054501},
  [\href{https://arxiv.org/abs/1506.07826}{{\ttfamily 1506.07826}}].

\bibitem{Yoon:2017qzo}
B.~Yoon, M.~Engelhardt, R.~Gupta, T.~Bhattacharya, J.~R. Green, B.~U. Musch
  et~al., \emph{{Nucleon Transverse Momentum-dependent Parton Distributions in
  Lattice QCD: Renormalization Patterns and Discretization Effects}},
  \href{http://dx.doi.org/10.1103/PhysRevD.96.094508}{\emph{Phys. Rev. D}
  {\bfseries 96} (2017) 094508},
  [\href{https://arxiv.org/abs/1706.03406}{{\ttfamily 1706.03406}}].

\bibitem{Zhang:2020dbb}
{\scshape Lattice Parton} collaboration, Q.-A. Zhang et~al., \emph{{Lattice QCD
  Calculations of Transverse-Momentum-Dependent Soft Function through
  Large-Momentum Effective Theory}},
  \href{http://dx.doi.org/10.1103/PhysRevLett.125.192001}{\emph{Phys. Rev.
  Lett.} {\bfseries 125} (2020) 192001},
  [\href{https://arxiv.org/abs/2005.14572}{{\ttfamily 2005.14572}}].

\bibitem{Shanahan:2020zxr}
P.~Shanahan, M.~Wagman and Y.~Zhao, \emph{{Collins-Soper kernel for TMD
  evolution from lattice QCD}},
  \href{http://dx.doi.org/10.1103/PhysRevD.102.014511}{\emph{Phys. Rev. D}
  {\bfseries 102} (2020) 014511},
  [\href{https://arxiv.org/abs/2003.06063}{{\ttfamily 2003.06063}}].

\bibitem{Luscher:2011kk}
M.~Luscher and S.~Schaefer, \emph{{Lattice QCD without topology barriers}},
  \href{http://dx.doi.org/10.1007/JHEP07(2011)036}{\emph{JHEP} {\bfseries 07}
  (2011) 036}, [\href{https://arxiv.org/abs/1105.4749}{{\ttfamily 1105.4749}}].

\bibitem{Chetyrkin:2003vi}
K.~Chetyrkin and A.~Grozin, \emph{{Three loop anomalous dimension of the heavy
  light quark current in HQET}},
  \href{http://dx.doi.org/10.1016/S0550-3213(03)00490-5}{\emph{Nucl. Phys. B}
  {\bfseries 666} (2003) 289--302},
  [\href{https://arxiv.org/abs/hep-ph/0303113}{{\ttfamily hep-ph/0303113}}].

\bibitem{Vladimirov:2017ksc}
A.~Vladimirov, \emph{{Structure of rapidity divergences in multi-parton
  scattering soft factors}},
  \href{http://dx.doi.org/10.1007/JHEP04(2018)045}{\emph{JHEP} {\bfseries 04}
  (2018) 045}, [\href{https://arxiv.org/abs/1707.07606}{{\ttfamily
  1707.07606}}].

\bibitem{Collins:2011zzd}
J.~Collins, \emph{{Foundations of perturbative QCD}}, vol.~32.
\newblock Cambridge University Press, 11, 2013.

\bibitem{GarciaEchevarria:2011rb}
M.~G. Echevarria, A.~Idilbi and I.~Scimemi, \emph{{Factorization Theorem For
  Drell-Yan At Low $q_T$ And Transverse Momentum Distributions
  On-The-Light-Cone}},
  \href{http://dx.doi.org/10.1007/JHEP07(2012)002}{\emph{JHEP} {\bfseries 07}
  (2012) 002}, [\href{https://arxiv.org/abs/1111.4996}{{\ttfamily 1111.4996}}].

\bibitem{Vladimirov:2016dll}
A.~A. Vladimirov, \emph{{Correspondence between Soft and Rapidity Anomalous
  Dimensions}},
  \href{http://dx.doi.org/10.1103/PhysRevLett.118.062001}{\emph{Phys. Rev.
  Lett.} {\bfseries 118} (2017) 062001},
  [\href{https://arxiv.org/abs/1610.05791}{{\ttfamily 1610.05791}}].

\bibitem{Echevarria:2012pw}
M.~G. Echevarria, A.~Idilbi, A.~Sch\"afer and I.~Scimemi,
  \emph{{Model-Independent Evolution of Transverse Momentum Dependent
  Distribution Functions (TMDs) at NNLL}},
  \href{http://dx.doi.org/10.1140/epjc/s10052-013-2636-y}{\emph{Eur. Phys. J.
  C} {\bfseries 73} (2013) 2636},
  [\href{https://arxiv.org/abs/1208.1281}{{\ttfamily 1208.1281}}].

\bibitem{Scimemi:2018xaf}
I.~Scimemi and A.~Vladimirov, \emph{{Systematic analysis of double-scale
  evolution}}, \href{http://dx.doi.org/10.1007/JHEP08(2018)003}{\emph{JHEP}
  {\bfseries 08} (2018) 003},
  [\href{https://arxiv.org/abs/1803.11089}{{\ttfamily 1803.11089}}].

\bibitem{Vladimirov:2019bfa}
A.~Vladimirov, \emph{{Pion-induced Drell-Yan processes within TMD
  factorization}}, \href{http://dx.doi.org/10.1007/JHEP10(2019)090}{\emph{JHEP}
  {\bfseries 10} (2019) 090},
  [\href{https://arxiv.org/abs/1907.10356}{{\ttfamily 1907.10356}}].

\bibitem{Bury:2020vhj}
M.~Bury, A.~Prokudin and A.~Vladimirov, \emph{{N$^3$LO extraction of the Sivers
  function from SIDIS, Drell-Yan, and $W^\pm/Z$ data}},
  \href{https://arxiv.org/abs/2012.05135}{{\ttfamily 2012.05135}}.

\bibitem{Mulders:1995dh}
P.~J. Mulders and R.~D. Tangerman, \emph{{The Complete tree level result up to
  order 1/Q for polarized deep inelastic leptoproduction}},
  \href{http://dx.doi.org/10.1016/0550-3213(95)00632-X}{\emph{Nucl. Phys. B}
  {\bfseries 461} (1996) 197--237},
  [\href{https://arxiv.org/abs/hep-ph/9510301}{{\ttfamily hep-ph/9510301}}].

\bibitem{Scimemi:2018mmi}
I.~Scimemi and A.~Vladimirov, \emph{{Matching of transverse momentum dependent
  distributions at twist-3}},
  \href{http://dx.doi.org/10.1140/epjc/s10052-018-6263-5}{\emph{Eur. Phys. J.
  C} {\bfseries 78} (2018) 802},
  [\href{https://arxiv.org/abs/1804.08148}{{\ttfamily 1804.08148}}].

\bibitem{EDWARDS2005832}
R.~G. Edwards and B.~Joó, \emph{The chroma software system for lattice qcd},
  \href{http://dx.doi.org/https://doi.org/10.1016/j.nuclphysbps.2004.11.254}{\emph{Nuclear
  Physics B - Proceedings Supplements} {\bfseries 140} (2005) 832--834}.

\bibitem{Bali:2016lva}
G.~S. Bali, B.~Lang, B.~U. Musch and A.~Sch\"afer, \emph{{Novel quark smearing
  for hadrons with high momenta in lattice QCD}},
  \href{http://dx.doi.org/10.1103/PhysRevD.93.094515}{\emph{Phys. Rev. D}
  {\bfseries 93} (2016) 094515},
  [\href{https://arxiv.org/abs/1602.05525}{{\ttfamily 1602.05525}}].

\bibitem{Hasenfratz:2001hp}
A.~Hasenfratz and F.~Knechtli, \emph{{Flavor symmetry and the static potential
  with hypercubic blocking}},
  \href{http://dx.doi.org/10.1103/PhysRevD.64.034504}{\emph{Phys. Rev. D}
  {\bfseries 64} (2001) 034504},
  [\href{https://arxiv.org/abs/hep-lat/0103029}{{\ttfamily hep-lat/0103029}}].

\bibitem{Maiani:1987by}
L.~Maiani, G.~Martinelli, M.~L. Paciello and B.~Taglienti, \emph{{Scalar
  Densities and Baryon Mass Differences in Lattice \{QCD\} With Wilson
  Fermions}}, \href{http://dx.doi.org/10.1016/0550-3213(87)90078-2}{\emph{Nucl.
  Phys. B} {\bfseries 293} (1987) 420}.

\bibitem{Bruno:2014jqa}
M.~Bruno et~al., \emph{{Simulation of QCD with N$_{f} =$ 2 $+$ 1 flavors of
  non-perturbatively improved Wilson fermions}},
  \href{http://dx.doi.org/10.1007/JHEP02(2015)043}{\emph{JHEP} {\bfseries 02}
  (2015) 043}, [\href{https://arxiv.org/abs/1411.3982}{{\ttfamily 1411.3982}}].

\bibitem{Engelhardt:2017miy}
M.~Engelhardt, \emph{{Quark orbital dynamics in the proton from Lattice QCD --
  from Ji to Jaffe-Manohar orbital angular momentum}},
  \href{http://dx.doi.org/10.1103/PhysRevD.95.094505}{\emph{Phys. Rev. D}
  {\bfseries 95} (2017) 094505},
  [\href{https://arxiv.org/abs/1701.01536}{{\ttfamily 1701.01536}}].

\bibitem{Engelhardt:2020qtg}
M.~Engelhardt, J.~R. Green, N.~Hasan, S.~Krieg, S.~Meinel, J.~Negele et~al.,
  \emph{{From Ji to Jaffe-Manohar orbital angular momentum in lattice QCD using
  a direct derivative method}},
  \href{http://dx.doi.org/10.1103/PhysRevD.102.074505}{\emph{Phys. Rev. D}
  {\bfseries 102} (2020) 074505},
  [\href{https://arxiv.org/abs/2008.03660}{{\ttfamily 2008.03660}}].

\bibitem{Shanahan:2019zcq}
P.~Shanahan, M.~L. Wagman and Y.~Zhao, \emph{{Nonperturbative renormalization
  of staple-shaped Wilson line operators in lattice QCD}},
  \href{http://dx.doi.org/10.1103/PhysRevD.101.074505}{\emph{Phys. Rev. D}
  {\bfseries 101} (2020) 074505},
  [\href{https://arxiv.org/abs/1911.00800}{{\ttfamily 1911.00800}}].

\bibitem{Moos:2020wvd}
V.~Moos and A.~Vladimirov, \emph{{Calculation of transverse momentum dependent
  distributions beyond the leading power}},
  \href{http://dx.doi.org/10.1007/JHEP12(2020)145}{\emph{JHEP} {\bfseries 12}
  (2020) 145}, [\href{https://arxiv.org/abs/2008.01744}{{\ttfamily
  2008.01744}}].

\bibitem{Qiu:1991pp}
J.-w. Qiu and G.~F. Sterman, \emph{{Single transverse spin asymmetries}},
  \href{http://dx.doi.org/10.1103/PhysRevLett.67.2264}{\emph{Phys. Rev. Lett.}
  {\bfseries 67} (1991) 2264--2267}.

\bibitem{Scimemi:2017etj}
I.~Scimemi and A.~Vladimirov, \emph{{Analysis of vector boson production within
  TMD factorization}},
  \href{http://dx.doi.org/10.1140/epjc/s10052-018-5557-y}{\emph{Eur. Phys. J.
  C} {\bfseries 78} (2018) 89},
  [\href{https://arxiv.org/abs/1706.01473}{{\ttfamily 1706.01473}}].

\bibitem{Schmelling:1994pz}
M.~Schmelling, \emph{{Averaging correlated data}},
  \href{http://dx.doi.org/10.1088/0031-8949/51/6/002}{\emph{Phys. Scripta}
  {\bfseries 51} (1995) 676--679}.

\bibitem{Ji:2019sxk}
X.~Ji, Y.~Liu and Y.-S. Liu, \emph{{TMD soft function from large-momentum
  effective theory}},
  \href{http://dx.doi.org/10.1016/j.nuclphysb.2020.115054}{\emph{Nucl. Phys. B}
  {\bfseries 955} (2020) 115054},
  [\href{https://arxiv.org/abs/1910.11415}{{\ttfamily 1910.11415}}].

\bibitem{Landry:2002ix}
F.~Landry, R.~Brock, P.~M. Nadolsky and C.~P. Yuan, \emph{{Tevatron Run-1 $Z$
  boson data and Collins-Soper-Sterman resummation formalism}},
  \href{http://dx.doi.org/10.1103/PhysRevD.67.073016}{\emph{Phys. Rev. D}
  {\bfseries 67} (2003) 073016},
  [\href{https://arxiv.org/abs/hep-ph/0212159}{{\ttfamily hep-ph/0212159}}].

\bibitem{Su:2014wpa}
P.~Sun, J.~Isaacson, C.~P. Yuan and F.~Yuan, \emph{{Nonperturbative functions
  for SIDIS and Drell\textendash{}Yan processes}},
  \href{http://dx.doi.org/10.1142/S0217751X18410063}{\emph{Int. J. Mod. Phys.
  A} {\bfseries 33} (2018) 1841006},
  [\href{https://arxiv.org/abs/1406.3073}{{\ttfamily 1406.3073}}].

\bibitem{Collins:2014jpa}
J.~Collins and T.~Rogers, \emph{{Understanding the large-distance behavior of
  transverse-momentum-dependent parton densities and the Collins-Soper
  evolution kernel}},
  \href{http://dx.doi.org/10.1103/PhysRevD.91.074020}{\emph{Phys. Rev. D}
  {\bfseries 91} (2015) 074020},
  [\href{https://arxiv.org/abs/1412.3820}{{\ttfamily 1412.3820}}].

\bibitem{Hunter:2007}
J.~D. Hunter, \emph{Matplotlib: A 2d graphics environment},
  \href{http://dx.doi.org/10.1109/MCSE.2007.55}{\emph{Computing in Science \&
  Engineering} {\bfseries 9} (2007) 90--95}.

\end{thebibliography}\endgroup
\end{document}